\documentclass[fleqn,usenatbib]{mnras}

\usepackage{newtxtext,newtxmath}

\usepackage[T1]{fontenc}
\usepackage{ae,aecompl}

\usepackage{graphicx}	
\usepackage{amssymb}	
\title[]{On the origin of central abundance drops in the intracluster medium of galaxy groups and clusters}
\author[A. Liu et al.]{
Ang Liu,$^{1,2,3}$\thanks{E-mail: liuang@arcetri.astro.it (A. Liu)}
Meng Zhai,$^{4}$\thanks{E-mail: mzhai@nao.cas.cn (M. Zhai)}
Paolo Tozzi$^{1}$
\\
$^{1}$INAF Osservatorio Astrofisico di Arcetri, Largo E. Fermi, I-50122 Firenze, Italy\\
$^{2}$Department of Physics, Sapienza University of Rome, I-00185 Rome, Italy\\
$^{3}$Department of Physics, University of Rome Tor Vergata, I-00133, Rome, Italy\\
$^{4}$Key Laboratory of Optical Astronomy, National Astronomical Observatories, Chinese Academy of Sciences, Beijing 100012, China\\
}

\date{Accepted XXX. Received YYY; in original form ZZZ}

\pubyear{2018}

\begin{document}


\maketitle

\begin{abstract}
A central drop of ICM Fe abundance has been observed in several cool-core clusters. It has been proposed
that this abundance drop may be due, at least partially, to the depletion of Fe into dust grains
in the central, high-density regions. According to this scenario, noble gas elements such as Ar and Ne
are not expected to be depleted into dust, and therefore should not show any drop, but follow the general
increase of metal abundance toward the center. In this work, we test this scenario by measuring
with {\sl Chandra} data the radial profiles of Ar and Ne in a sample of 12 groups and clusters where a
central drop in Fe abundance has been detected. We confirm the presence of the Fe drop in 10 out of 12
clusters at more than 2$\sigma$ c.l., and 4 Ar drops with similar significance. We also find 4 Ne
drops, with the caveat that Ne abundance measurement from CCD spectra suffers from systematics not
completely under control. Our results are consistent with an abundance drop common to the three elements. When comparing the profiles, we find that, on average, the abundance
profiles of Ar and Ne are significantly higher than Fe and steeper toward the center, while they all
gradually become consistent with solar composition at $r\geq 0.05r_{500}$. We also check that
Si and S profiles are mostly consistent with Fe. This result confirms a scenario in
which some fraction of Fe is depleted into dust grains in the inner regions, although the global
central abundance drop is mostly due to mechanical processes, like the displacement of metal-rich ICM from
the very center to larger radii by AGN-driven feedback.  Finally, we report the detection of an Fe drop in
the cluster MACSJ1423.8+2404 at $z=0.543$, showing that this feature appears early on in
cool-core clusters.
\end{abstract}



\section{Introduction}

X-ray observations of galaxy groups and clusters have shown that the intracluster medium (ICM)
in the core can be so dense that the central cooling time is significantly shorter than
the cluster lifetime, particularly in the innermost few tens of kpc, as measured in high
resolution {\sl Chandra} data \citep[see][]{2009Cavagnolo}.
If the cooling proceeded isobarically, the ICM would be doomed to cool rapidly,
forming a cooling flow with a mass deposition rate $\dot M$ that can reach $100-1000 M_\odot$ yr$^{-1}$, as
directly derived from the total X-ray luminosity of the core region \citep[see][]{fabian1994}.
Among the unavoidable consequences, one would expect a large amount of cold gas falling
onto the central galaxy (called the brightest cluster galaxy, from now on BCG)
and a star formation rate (SFR) comparable to the predicted $\dot M$. However, the
observed SFRs in BCGs are observed to be modest, of the order of few $M_\odot$ yr$^{-1}$
(except few relevant cases),
consistent with being entirely fueled by the stellar mass loss from the BCG itself \citep{voit2011}.
In addition, the amount of cold gas has been known since long time to be much lower than that
expected from the high cooling-flow rates, thanks to the measurements of CO emission from
molecular gas and other observables \citep[see][]{mcnamara1989,edge2001}. The final blow to the
isobaric cooling-flow
paradigm was provided to the observations with the Reflection Grating Spectrometer (RGS) onboard
XMM-Newton, whose high resolution spectra of cluster cores showed the absence of the metal emission lines
associated to gas cooler than $\sim 1/3$ of the virial temperature \citep{peterson2001,peterson2006}.
This finding, in turn, implies that the bulk of the gas in the core must be prevented from cooling by
some heating mechanisms, leaving only a minority of the ICM, if any, available for complete cooling.
Eventually, the cooling-flow model has been superseded by the cool-core picture \citep{molendi2001}.

Many candidates have been proposed as contributors of the heating process.  Among them: cosmic rays
\citep{bohringer1988}, star formation and supernovae \citep{domainko2004,li2012},
dissipation of gas turbulence induced either by buoyancy of bubbles inflated by the central active
galactic nucleus (AGN) or by minor and major mergers
\citep[see][]{mcnamara2007,fabian2012,kravtsov2005,heinz2006,2013Gaspari},
or a combination of mixing and heating associated to vortices formed during the inflation of the bubbles
\citep{hillel2016}. The large amount of mechanical energy stored in the bubbles carved in the ICM by
radio jets makes AGN feedback the strongest candidate as the main heating source.
Thanks to the unprecedented angular resolution of {\sl Chandra}, the detailed X-ray image of the
Perseus cluster presented for the first time a clear view of the complex structures arising
in the ICM due to AGN feedback, such as cavities and ripples \citep{fabian2003,fabian2006}.  Recently,
the observations of the Perseus cluster with {\sl Hitomi} provided a unique measurement of the ICM dynamics
associated to the feedback process \citep{hitomi2016nat}.  This increasing amount of evidence clearly shows
that the dynamics and thermodynamics of the ICM are strongly affected by AGN feedback, although the
discussion on the heating process is still open, and we are far from having a coherent picture of the
cycle of baryons in cool cores.

An unambiguous signature of these processes is left in the metal abundance distribution
and its evolution, a topic which rapidly gained attention from the scientific community since the beginning
of the era of {\sl Chandra} and XMM-Newton. Evidence of chemo-dynamical impact of AGN feedback are now
directly observed in the X-ray data \citep{simionescu2009,kirkpatrick2009,osullivan2011,lovisari2019}, and, at the same
time, supported by continuously improving hydrodynamical simulations \citep{roediger2007,sijacki2007,churazov2013}.
A striking example is provided by the observed enhancement of metals around the
bubbles \citep{kirkpatrick2011}, strongly suggesting that the inflation of the cavities uplifts
the highly enriched ICM bringing it at larger radii. Further insights came from detailed studies of the
metallicity peaks commonly observed at the center of cool-core clusters at moderate and high redshift
\citep[][]{degrandi2004,2018Mcdonald}.  In particular, the measurement of the broadening of the distribution
of metals across cosmic epochs in cool-core clusters provides further support to the nuclear activity as the
main contributor of metal mixing in the ICM \citep{degrandi2014}. Recently, in \citet{liu2018} we showed that
the width of the iron (Fe) peak increases by a factor of $\sim$3 from redshift $z\sim 1$ to $z \sim 0.05$.

In this paper we focus on an apparently minor property of the metal distribution in the ICM,
which, instead, may provide a relevant missing link between the hot and cold phases of the
baryons cycling in and around the BCGs.  It consists in a small-scale (of the order of 10 kpc) decrement
at the center of the abundance peak in cool cores, a feature that we call {\sl central abundance drop},
and that has long been known thanks to {\sl Chandra} and XMM-Newton studies of single, low-redshift
clusters \citep[][]{sanders2002,johnstone2002,sanders2007,million2010,panagoulia2013} and confirmed
recently by more extensive studies \citep{panagoulia2015,mernier2017}.
This feature is better observed exploiting the high
spatial resolution of the {\sl Chandra} satellite in cool-core clusters, but has also been confirmed with
XMM-Newton in low redshift clusters \citep{mernier2017}.  In addition, the abundance drop has been
observed both in Fe, which is the most prominent line-emitting element in X-ray spectra, and in other elements
like sulphur (S) and silicon (Si) \citep{mernier2017,panagoulia2015}.  The Fe abundance drop has been
observed usually in association with cavities, and therefore it may be ascribed to the
local effects of AGN feedback, effectively pushing the highly-enriched gas from the innermost regions
toward larger radii \citep{simionescu2009,kirkpatrick2011}.  As a result, the measured
abundance will decrease somehow at small radii and correspondingly increase at larger radii,
at the point of inverting locally the abundance gradient on a spatial scale corresponding to the
efficiency of the uplifting of the metal-enriched gas, and creating the abundance drop.
Clearly, the uplifting of the innermost, mostly enriched gas implies a similar
decrease for all the elements, also if the mechanism is not due to the inflation of bubbles from the ICM,
but to other mechanical processes like core sloshing \citep{ghizzardi2014}.

Other possible interpretations have been proposed, like contamination from a central X-ray emitting
AGN in the ICM spectrum, or an underestimate of the helium content at small radii as a result of mass
sedimentation \citep{ettori2006,mernier2017}.  However, the former interpretation cannot explain
abundance drops which are observed far from the AGN position, nor the second interpretation can justify
why many cool-core clusters do not exhibit central abundance drop. Another possible effect is the
non-negligible optical depth of some resonance line, that may result, in principle, in
a decrease of the global ion abundance measured assuming an optically-thin medium in the densest ICM regions.
Resonance scattering has been now measured for the Helium-like iron emission line by {\sl Hitomi}
\citep{2018_Hitomi_RS}, however, it appears to be relevant only in the analysis of high-resolution spectra,
while it has a limited impact on the global metallicity measured from CCD spectra and it is not effective
in removing the central dip \citep[see][]{2006Sanders,2017Gendron}. See, however \citet{simionescu2019}
for a more comprehensive discussion on the comparison of high-resolution spectra with those obtained
from CCD data on the measurement of the abundance pattern in the ICM.

Finally, a relevant physical mechanism that may play a role here, is the depletion of Fe and several other
metals (including Si and S) into dust grains present in the diffuse gas close to the BCG.
This is supported by the detection of cold dust filaments in optical and infrared in clusters with central
abundance drop \citep{panagoulia2013,panagoulia2015}. Depletion of Fe and other reactive metals
is expected if the enriched gas from star formation and stellar mass-loss in the BCG remains cool and distinct
from the hot surrounding ICM, due to the higher ambient pressure \citep{voit2011}.  Being trapped into
dust filaments, a significant part of Fe is released into the ICM on a longer time scale and
displaced with respect
to non-reactive elements, namely when the dust-rich gas is uplifted by nuclear activity into lower-density
and hotter regions \citep[see discussion in][]{panagoulia2013}. This dust-depletion mechanism can be
tested by comparing the radial abundance profiles of Fe to that of noble gas elements that are not expected
to be embedded into dust grains.  The two noble elements that are potentially detectable in high S/N spectra
of clusters are neon (Ne) and argon (Ar). While the former is hard to measure because blended with the
dominant emission from the Fe L-shell complex in CCD spectra, the latter can be firmly detected thanks to
the Ar XVII and Ar XVIII emission lines which appear isolated in the range 3.0--3.3 keV for low-$z$ clusters.
Therefore, in this paper we aim at exploiting the angular resolution of {\sl Chandra} data to measure
the spatially-resolved abundance profiles of Ar, and possibly of Ne, in a sample of groups and
clusters in which a central Fe abundance drop has been previously detected, and compare them to
the profiles of Fe and other elements.

The paper is organized as follows. In Section 2, we describe the sample of clusters and groups with previous
detection of central abundance drops, the data we use in this work, and our analysis strategy. In Section 3,
we present the measured radial abundance profiles of Fe, Ar, Ne, Si and S, and show how our results
can be used to constrain the dust grain scenario. In Section 4, we discuss the reliability
of our spectral analysis strategy. Our conclusions are summarized in Section 5. Throughout this paper,
we adopt the seven-year WMAP cosmology with $\Omega_{\Lambda} =0.73 $, $\Omega_m =0.27$, and $H_0 = 70.4$
km s$^{-1}$ Mpc$^{-1}$ \citep{2011Komatsu}. Quoted error bars correspond to a 1 $\sigma$ confidence level,
unless noted otherwise.

\section{Cluster Sample, Data reduction and analysis}

\begin{table*}
\caption{The sample of groups and clusters with a significant detection of central Fe abundance drop
considered for this work.  The exposure times are computed after data reduction. The values of $r_{500}$
are taken from the cluster compilation in \citet{pinto2015} and \citet{liu2018}.
The central Fe abundance drop detection is obtained from the following references:
P15: \citet{panagoulia2015}; M17: \citet{mernier2017}; SF02: \citet{sanders2002}; M10: \citet{million2010};
 J02: \citet{johnstone2002}; L18: \citet{liu2018}. }
 \begin{tabular}{lcclcl}
\hline
Cluster name   & $z$ & $r_{500}$(Mpc) & {\sl Chandra} ObsID & Exptime(ks)  & Reference  \\
\hline
NGC4636          & 0.0031 & 0.35 & 323,324,3926,4415        & 203.5  & P15, M17   \\
NGC5846          & 0.0057 & 0.36 & 7923                     & 90.0  & P15, M17   \\
NGC5813          & 0.0065 & 0.44 &  5907,9517,12951--3,13246-7,13253,13255   & 635.3  & P15, M17   \\
NGC5044          & 0.0093 & 0.56 & 798,9399,17195--6,17653,17654,17666     & 417.7  & P15, M17    \\
Centaurus        & 0.0114 & 0.83 & 4954--5,5310,16223--5,16534,16607--10     & 666.3  & SF02, P15, M17   \\
A262             & 0.0161 & 0.74 &  2215,7921                & 28.7  & P15 \\
A3581            & 0.0220 & 0.72 & 12884                    & 83.9  & M17    \\
Ophiuchus        & 0.0280 & 1.22 & 16142--3,16464,16626--7,16633--5,16645,3200 & 259.9 & M10    \\
A2199            & 0.0303 & 1.00 & 497--8,10748,10803--5    & 157.3  & J02  \\
2A0335           & 0.0349 & 1.05 & 919,7939,9792            & 102.5  & P15, M17    \\
A1991            & 0.0590 & 0.82 & 3193                     & 38.1  & P15  \\
MACSJ1423.8+2404 & 0.5431 & 0.97 & 4195                     & 38.9  & L18     \\
  \hline
 \end{tabular}
\label{sample}
\end{table*}

\subsection{Selection of the sample}

Our main scientific goal is to find a signature of dust depletion in the ICM of galaxy groups and clusters,
in terms of a selective decrement in the abundance of Fe, Si and S with respect to noble
elements in the central regions.
Dust depletion into dust grains is expected to be efficient in high density environments, where some of
the enriched gas, particularly that associated to the stellar mass-loss, remains cool and distinct
from the hot ICM due to the high pressure. Therefore, to constrain the impact of dust depletion on the
Fe distribution, we select a sample of cool-core clusters with a firm detection of a
central Fe drop. Due to the high angular resolution needed for this study, we also require the
use of {\sl Chandra} data.

So far, the largest systematic investigation of abundance drop across the {\sl Chandra} archive is provided
by the work of \citet{panagoulia2015}, where eight groups and clusters are reported to have central abundance
drop with high statistical significance (NGC 4696/Centaurus, NGC 5846, 2A0335+096, NGC 4636 Abell 1991,
Abell 262, NGC 5813, NGC 5044). Detection of Fe drop in single objects are reported for Abell 2199
\citep{johnstone2002}, Ophiuchus \citep{million2010}, NGG 4696/Centaurus \citep{sanders2002,panagoulia2013},
and Perseus \citep{sanders2007}. In addition, 14 objects have been shown to have a decrease of the Fe
abundance in their core on the basis of XMM-Newton data in \citet{mernier2017}.
All of them have been observed also by {\sl Chandra}, and six were already included among the clusters with
significant drop in \citet{panagoulia2015}.  Among the remaining eight, one is the Perseus clusters,
while the remaining seven are M49, M86, Abell 189 (NGC 533), Fornax, HCG62, NGC4325 and Abell 3581,
with the last three also included in \citet{panagoulia2015} among those with low-significance Fe drop.

To draw a final list of targets relevant for our study, we combine all these sources and perform a
preliminary test
on the detectability of the Ar line complex, by fitting the spectrum of each cluster in the innermost
20 kpc, then setting the Ar abundance $Z_{\rm Ar}$ to zero and checking whether the $\Delta C_{\rm stat}
\ge 2.7$. With this criterion, we select all the 8 clusters with high-significance drop in
\citet{panagoulia2015}, plus Abell 2199, Ophiuchus, and Abell 3581. We exclude the Perseus cluster
\citep{sanders2007} since the exceptional data quality for this cluster (1.4~Ms total exposure with
{\sl Chandra}) allows one to attempt such an investigation in single 2D regions, rather than an
azimuthally-averaged profile.  For this reason we decided to postpone the analysis of Perseus to another
paper with a dedicated 2D, spatially-resolved spectral analysis.

Therefore, we identify 11 groups and clusters viable for our study.  Since all these well-documented cases
are found at low redshift ($z<0.1$), we also include, for the sake of discussion, a high-redshift
cluster where we serendipitously detected for the first time a shallow central drop
\citep[MACSJ1423.8+2404 at $z=0.543$]{liu2018}.  The list of the 12 clusters and groups discussed in
this work is shown in Table \ref{sample}, where we also report the X-ray data, redshift, total exposure
time and $r_{500}$ as measured in the literature. Clearly, we made best use of the
{\sl Chandra} archive, and include all the available observations for each cluster.  However, we exclude
a few obsid when they have different data acquisition mode or period, and, at the same time, provide
a minor contribution to the total exposure time.  Therefore, we discard Obsid 788 for NGC5846, Obsid 1650
for A3581, Obsid 1657 for MACSJ1423, and Obsid 504, 505, 4190 and 4191 for Centaurus.
We remind that we are not interested in a complete sample
of clusters and groups with a central abundance drop, but we just require a reliable sample to investigate
whether we can identify the signature of dust depletion with our analysis strategy. Essentially, our
sample is mostly based on the results of \citet{panagoulia2015} and \citet{mernier2017} for the Fe profile,
with the inclusion of a few other well studied cases.

\subsection{Data reduction}

Data reduction is performed with {\tt CIAO 4.10}, adopting the latest release of the
{\sl Chandra} Calibration Database at the time of writing
{\tt (CALDB 4.7.8)}. Unresolved sources within the ICM, typically due to AGN not associated to the cluster,
are identified with {\tt wavdetect}, checked visually and eventually removed. In particular, we pay attention
to possible nuclear emission from the BCG, that may affect the spectrum of the innermost bin.
Nuclear emission from the BCG is found in about 20\% of all the clusters, and preferentially in cool cores,
despite the AGN luminosity is typically low, of the order of $10^{43}$ erg s$^{-1}$ \citep[see][]{2018Yang}.
Thanks to the angular resolution of {\sl Chandra}, this consists in removing a circle of $\sim 1.2$ arcsec
at the center of the innermost bin.

The spectra are extracted in circular concentric rings, centered at the peak of X-ray
emission in 0.5--7 keV band image, smoothed with a Gaussian function with full width at half-maximum
$\sim 1.5$ arcsec after removing unresolved sources. The annuli are chosen adaptively on the image to
ensure a roughly equal number of net counts, and in any case larger than 2000 in the 0.5--7 keV band,
in each annulus. For clusters with multiple observations, we extract the spectrum and compute the response
matrix file (RMF) and ancillary response file (ARF) for each observation separately. Background spectra are
extracted from regions far from the ICM emission in each Obsid. When the ICM emission
fills the entire CCD, we use the background generated from the
`blank sky' files with the {\tt blanksky} script.  This choice may not be accurate in the outermost regions
beyond $0.1r_{500}$ which, however, are not relevant in this work.  In the core regions
we discuss here, the accuracy of few percent provided by the `blank sky' background spectra
is not affecting our results.

\subsection{Spectral analysis strategy}

The spectral fits are performed with {\tt Xspec 12.10.1} \citep{1996Arnaud} using C-statistics
\citep{cash1979}.  All the abundance values in this paper are relative to the solar values of
\citet{asplund2009}. We note that these photospheric results are significantly smaller, for the most
abundant elements like C, N, O, Ne and Fe, than those recommended in the widely used compilations of
\citet{1989Anders} and \citet{1998Grevesse}.  In particular, the Fe abundance in solar units turns out to
be larger by a factor $\sim 1.6$.  Galactic absorption is described by the model {\tt tbabs}
\citep{2000Wilms}, where the best-choice for the value of Galactic HI column density is initially set to
the value corresponding to the cluster position in the HI survey of \citet{2005Kalberla}. However, we
notice in several cases that a different value of $NH_{Gal}$ is needed to obtain an acceptable fit,
therefore we leave the $NH_{Gal}$ parameter free to vary (see the Discussion section for further details).
For clusters with multiple observations, the spectra corresponding to the same ring are fitted with linked
parameters.

Particular care must be taken in modeling the thermal structure of the ICM, since we necessarily have more
than one temperature component in each spectrum.  This may be simply due to projection effects
of background and foreground ICM belonging to outer shells in presence of strong temperature and
abundance gradients across the line of sight, particularly in the innermost regions. We also know that the
complex ICM structure in strong cool-cores with cavities can create strong departure from spherical
symmetry. This complexity is clearly emerging thanks to the high signal spectra in our sample,
and, therefore, it is not possible to assume a dominant single-temperature component in each bin,
particularly because this choice may bias the metallicity low, as noticed in several studies
\citep{2000Buote,2003Buote,2010Gastaldello}. Therefore, in this work we adopt the simplest approach
which provides statistically acceptable fits, which consists in using two temperatures in each spectrum.
Previous works investigating the abundance drop in cool core clusters already showed that a
two-temperature model is relatively unaffected by projection effects and return reliable values of
metal abundance \citep{panagoulia2013}. We also verify {\sl a posteriori} the goodness of the fit to
make sure that our strategy provides an acceptable modelization of the observed spectrum.
Finally, we assume that the abundance of each element we consider is the same in both temperature
components, a choice that is necessary to obtain robust results and will be further detailed in the
Discussion section.

The emission lines of Ar and Ne have very different diagnostics.  The signal from Ar is contributed
by the $K_\alpha$ lines of Ar XVII and Ar XVIII (He-like and H-like Ar) which appear clearly above the continuum
in the energy range $\sim$3.0--3.3~keV,
far from other emission lines.
In this case, the capability of detecting Ar depends merely on the strength of the signal, on the Ar
abundance itself, and on the ICM temperature, so that the associated error is essentially dominated by
Poisson noise.  The diagnostics for Ne is completely different, since the $K_\alpha$ lines from
Ne X and Ne IX are blended with the overwhelming Fe XXII--XIX line complex (plus the less prominent
O VII and O VIII line complex) at $\sim 1$~keV, and cannot be resolved into single lines at the spectral
resolution of ACIS-I or ACIS-S. Clearly, the value and the associated error of the Ne abundance is
dramatically dependent on the capability of properly modeling the complex line blend around 1~keV, which may
well not reproduced with only two temperatures. For this reason the Ne profile has been measured only in
a few cases so far \citep[see][]{2004dePlaa}. Despite that, we present and discuss our measurements of Ne
abundance obtained with the two-temperature assumption, with the caveat that the associated errors is
a lower limit to the real uncertainty.

According to this scheme, we measure the abundance of metals by fitting the projected spectrum in each
annulus with a double {\tt vapec} thermal plasma emission model \citep{smith2001} with AtomDB
version 3.0.9, leaving free the two temperatures and the corresponding emission measure (normalization),
the Galactic absorption $NH_{Gal}$, and the abundances of Fe, Ar, Ne, S, Si, O, Mg, and Ca which
are assumed to be the same in the two temperature components. All the other abundance values are linked to Fe.
The abundance of Helium is always fixed to solar value.

Finally, in this work we focus on the projected quantities, with no attempt to deproject the
abundance profiles we obtain, similarly to the approach of \citet{mernier2017}. Deprojection procedures, which
are commonly adopted in the literature mostly to obtain 3D temperature profiles and measure hydrostatic masses
\citep[see, e.g.,][]{churazov2003,2008Russell} are known to have little effects on the abundance profiles
\citep[see also][]{2007Rasmussen}.  In addition, the complexity of the thermal structure
in cool cores, particularly with evident cavities, would result in an incorrect result if deprojected
under the assumption of spherical symmetry, as noticed by \citet{panagoulia2015}.

\section{Radial Abundance Profiles}

\begin{table*}
\centering
\caption{The magnitude of the abundance drop, defined as $\Delta Z = Z_{\rm peak}-Z_{\rm in}$, for
Fe, Ar and Ne is shown in column 2, 3, and 4, respectively.  In column 5, 6, and 7, we list the
distance between the maximum of the X-ray emission and the peak of the abundance profiles in kpc.
The offset between the BCG and the X-ray emission peak is also given in column 8. }
 \begin{tabular}{lccccccc}
\hline
Cluster   & $\Delta Z$ (Fe) & $\Delta Z$ (Ar) & $\Delta Z$ (Ne) & $D_{\rm Fe}$ (kpc)  & $D_{\rm Ar}$ (kpc)  & $D_{\rm Ne}$ (kpc) & $D_{\rm BCG}$ (kpc) \\
\hline
NGC4636          & $ 0.25\pm0.08$ & $ 0.86\pm0.92 $ & -               & $2.9\pm 2.0   $ & $1.3	\pm0.7	$ & -                 & 0.16 \\
NGC5846          & $ 0.49\pm0.17$ & $ 1.78\pm0.95 $ & $ 1.04\pm0.45 $ & $4.3\pm 2.2   $ & $6.4	\pm3.4 	$ & $ 4.4	\pm 1.7	 $  & 0.12 \\
NGC5813          & $ 0.71\pm0.05$ & $ 0.92\pm0.45 $ & $ 1.06\pm0.30 $ & $14.8\pm2.5   $ & $12.2\pm4.3	$ & $ 14.0\pm 5.4$    & 0.16 \\
NGC5044          & $ 0.09\pm0.04$ & -             & $ 0.26\pm0.26 $ & $5.5\pm 2.5   $ & -   & $ 10.4\pm 7.1	 $  & 0.22 \\
Centaurus        & $ 0.51\pm0.12$ & $ 0.58\pm0.16 $ & -               & $14.9\pm4.8	  $ & $10.4\pm5.6	$ & -	                & 1.56 \\
A262	           & $ 0.74\pm0.10$ & $ 1.22\pm0.90 $ & $ 1.11\pm0.74 $ & $14.3\pm5.4	  $ & $10.7\pm6.0	$ & $ 15.4\pm 9.0 $   & 0.99 \\
A3581            & $ 0.24\pm0.08$ & $ 0.17\pm0.74$  & $ 1.01\pm0.45 $ & $33.1\pm12.6	$ & $ 21.1\pm 21.1 $  & $ 19.0\pm 11.0 $  & 0.19 \\
Ophiuchus        & $ 0.42\pm0.16$ & $ 1.03\pm 1.29$ & -               & $3.5 \pm 0.7  $ & $3.1 \pm1.5$   & -                 & 2.25 \\
A2199            & $ 0.07\pm0.07$ & $ 1.18\pm0.56 $ & $ 0.63\pm0.42 $ & $5.6\pm 3.1	  $ & $7.8	\pm3.1	$ & $ 8.5	\pm 3.9	 $  & 1.98 \\
2A0335           & $ 0.03\pm0.09$ & $ 0.41\pm0.40 $ & -               & $46.7\pm42.7	$ & $56.7\pm33.5	$ & -                 & 7.06 \\
A1991	           & $ 0.57\pm0.16$ & $ 0.99\pm0.45 $ & $ 1.25\pm0.61 $ & $27.5\pm10.0	$ & $21.9\pm10.4 $ & $ 17.8\pm 7.3	 $  & 4.76 \\
MACSJ1423.8+2404 & $ 0.25\pm0.12$ & $ 1.15\pm0.76$   & $ 1.63\pm1.45 $ & $20.4\pm7.7	  $ & $ 24.7\pm 11.0 $ & $ 24.3\pm 8.3 $   & -    \\
\hline
\end{tabular}
\label{result}
\end{table*}

In this work we are mostly interested in the comparison of the Fe profiles to those of Ar and, with
the mentioned caveats, Ne.  We also discuss S and Si that are expected to be also depleted similarly to
Fe, while we are not able to obtain an accurate characterization of the profiles of O, Ca and Mg,
which, therefore, are not included in this investigation.

\begin{figure*}
\begin{center}
\includegraphics[width=3.4in, height=1.45in, trim=80 410 80 200, clip]{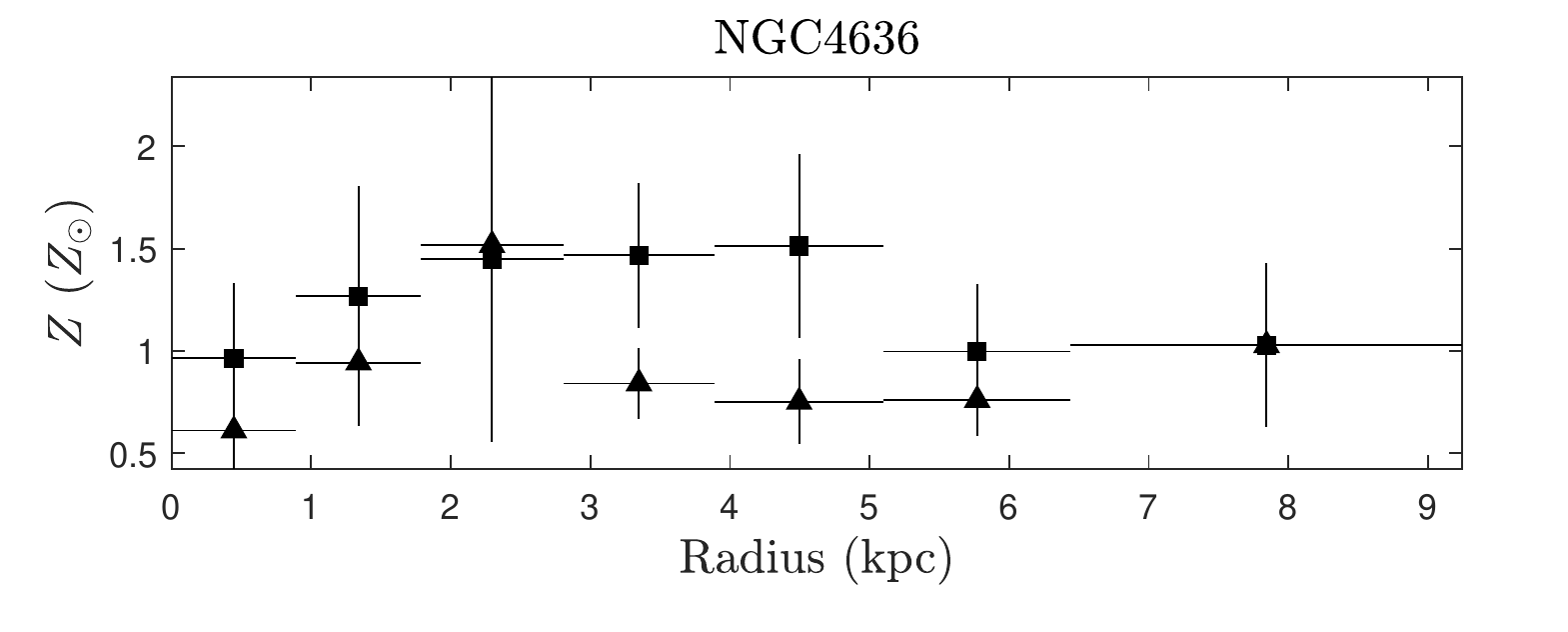}
\includegraphics[width=3.4in, height=1.45in, trim=80 410 80 200, clip]{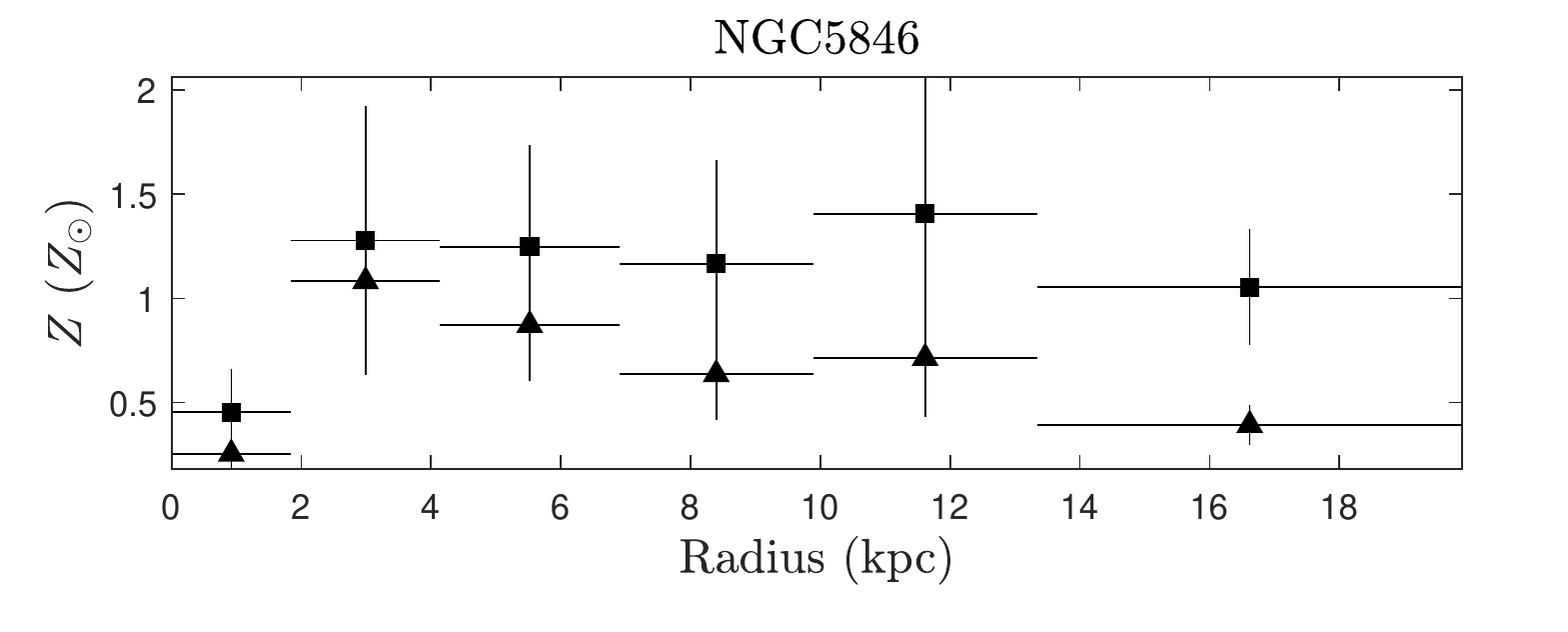}
\includegraphics[width=3.4in, height=1.45in, trim=80 410 80 200, clip]{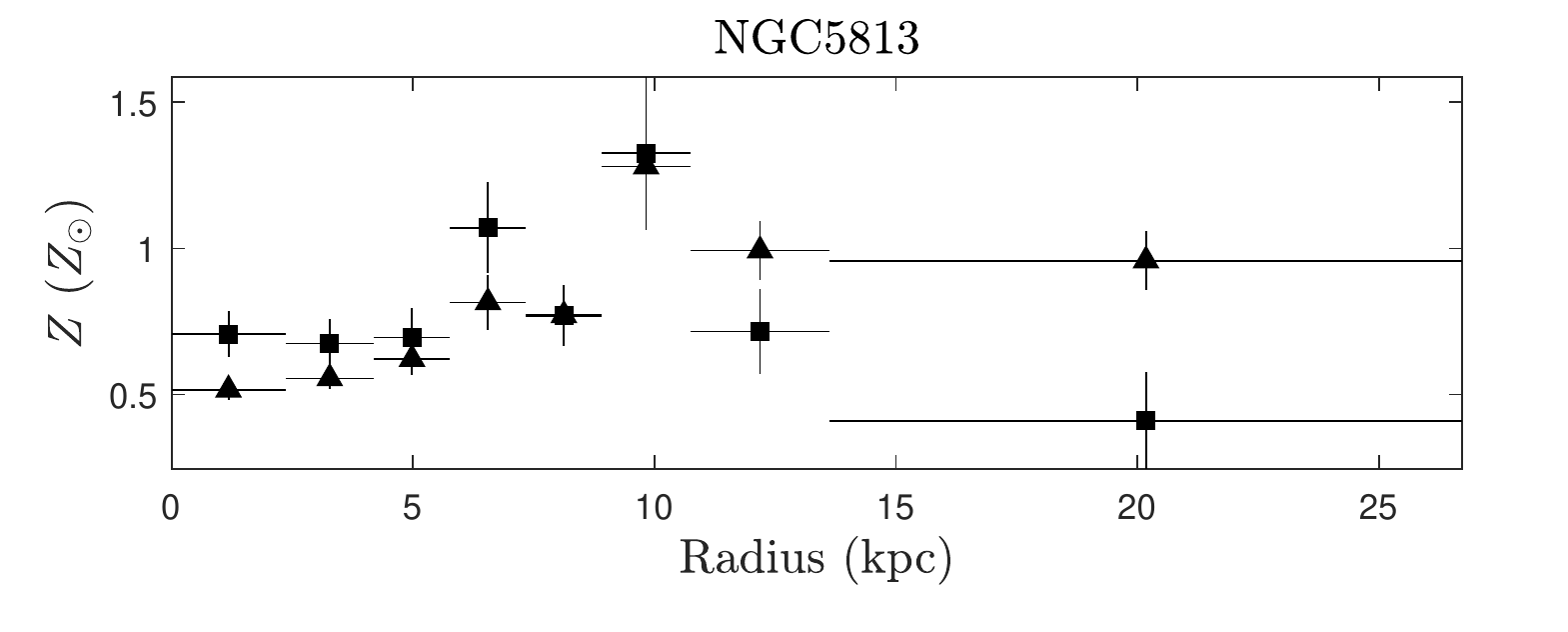}
\includegraphics[width=3.4in, height=1.45in, trim=80 410 80 200, clip]{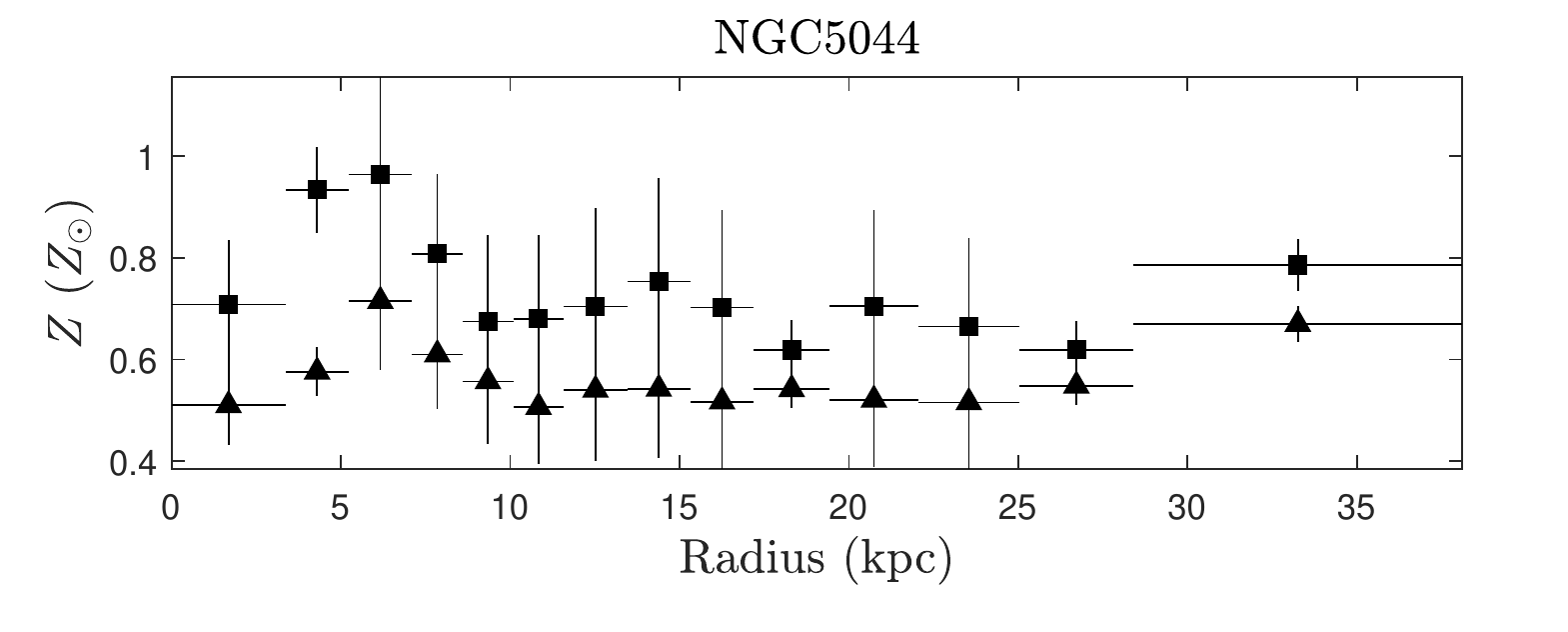}
\includegraphics[width=3.4in, height=1.45in, trim=80 410 80 200, clip]{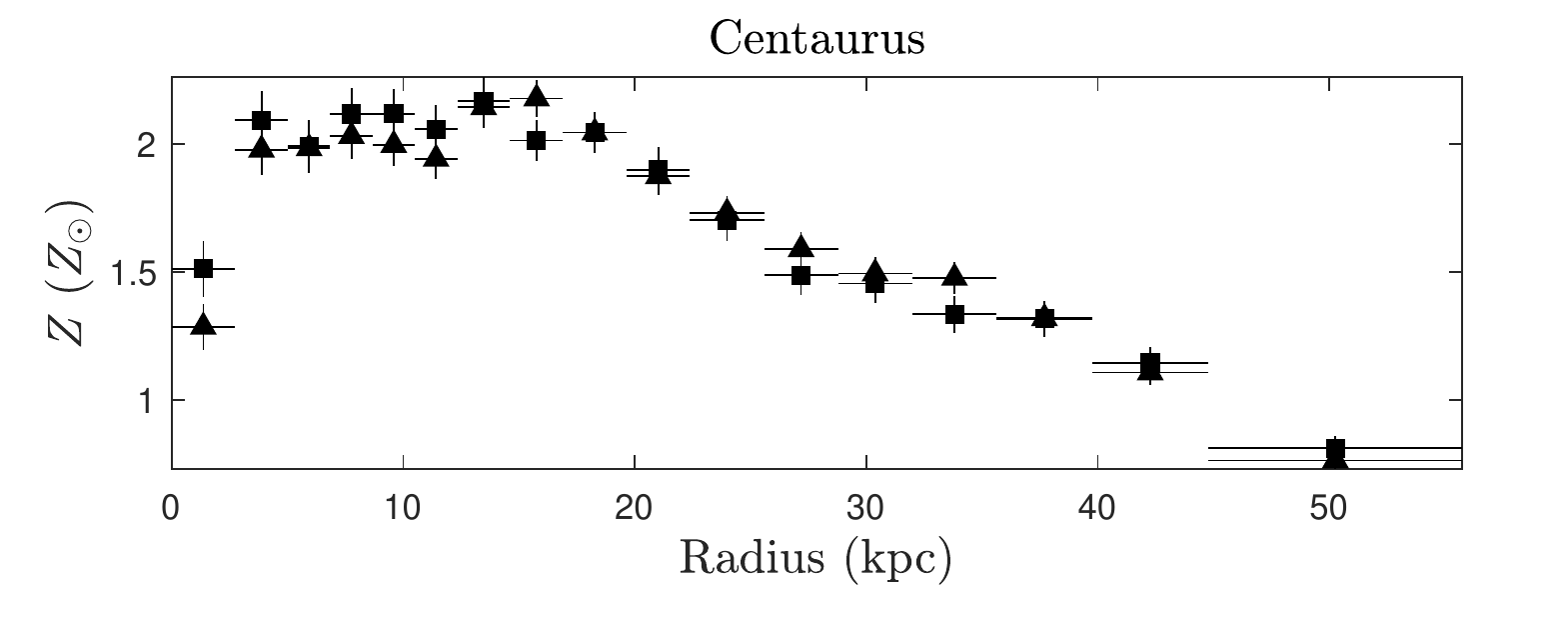}
\includegraphics[width=3.4in, height=1.45in, trim=80 410 80 200, clip]{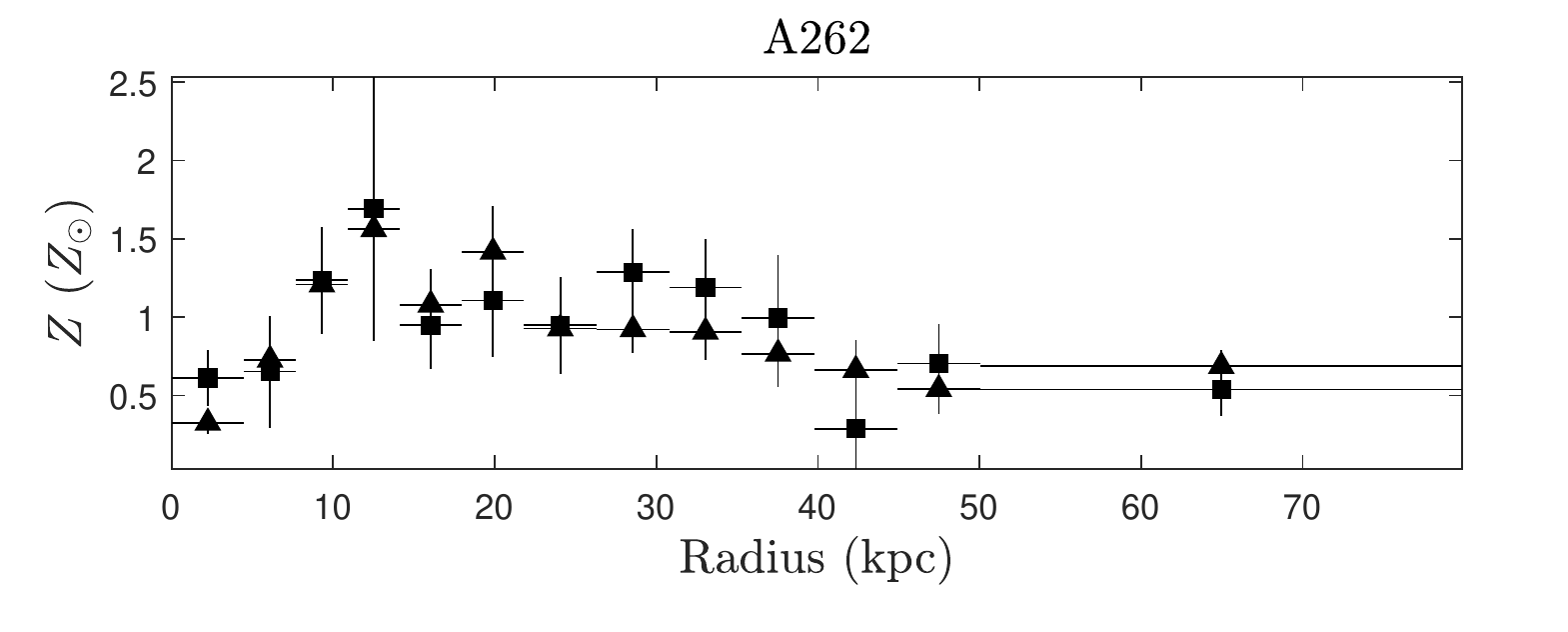}
\includegraphics[width=3.4in, height=1.45in, trim=80 410 80 200, clip]{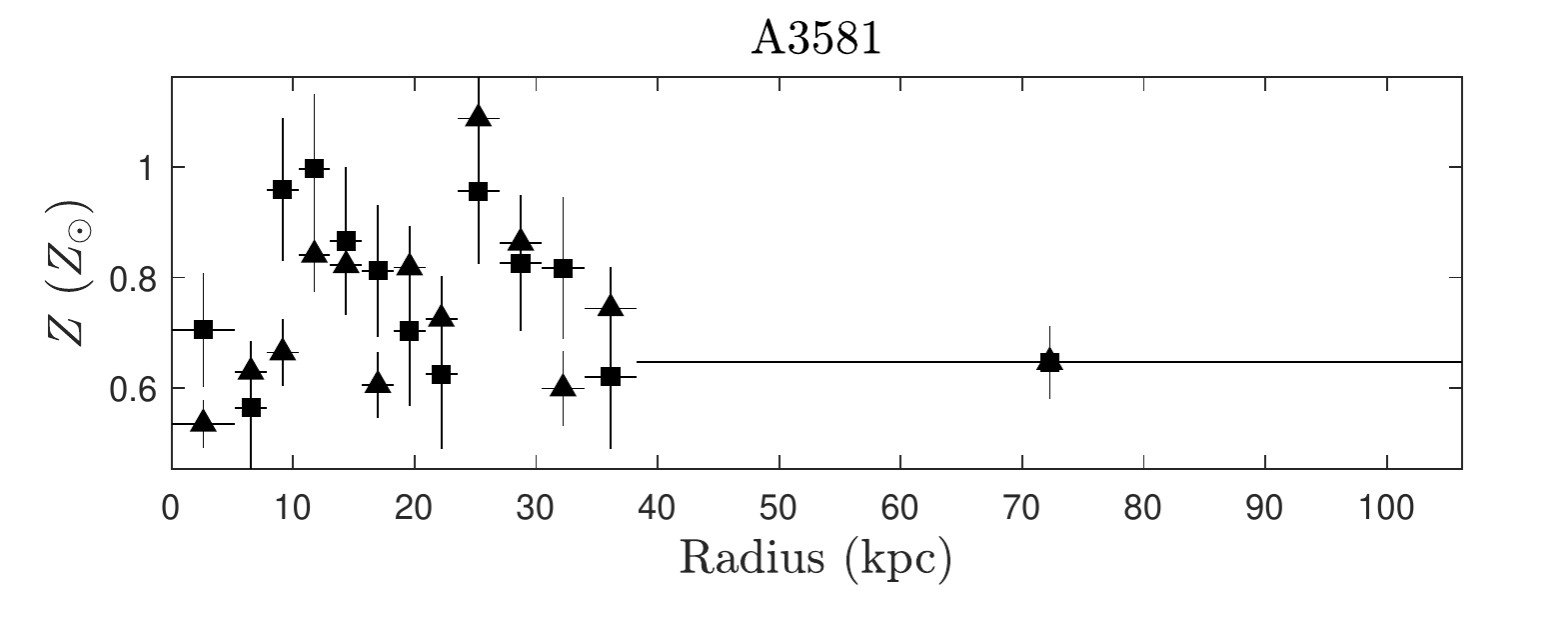}
\includegraphics[width=3.4in, height=1.45in, trim=80 410 80 200, clip]{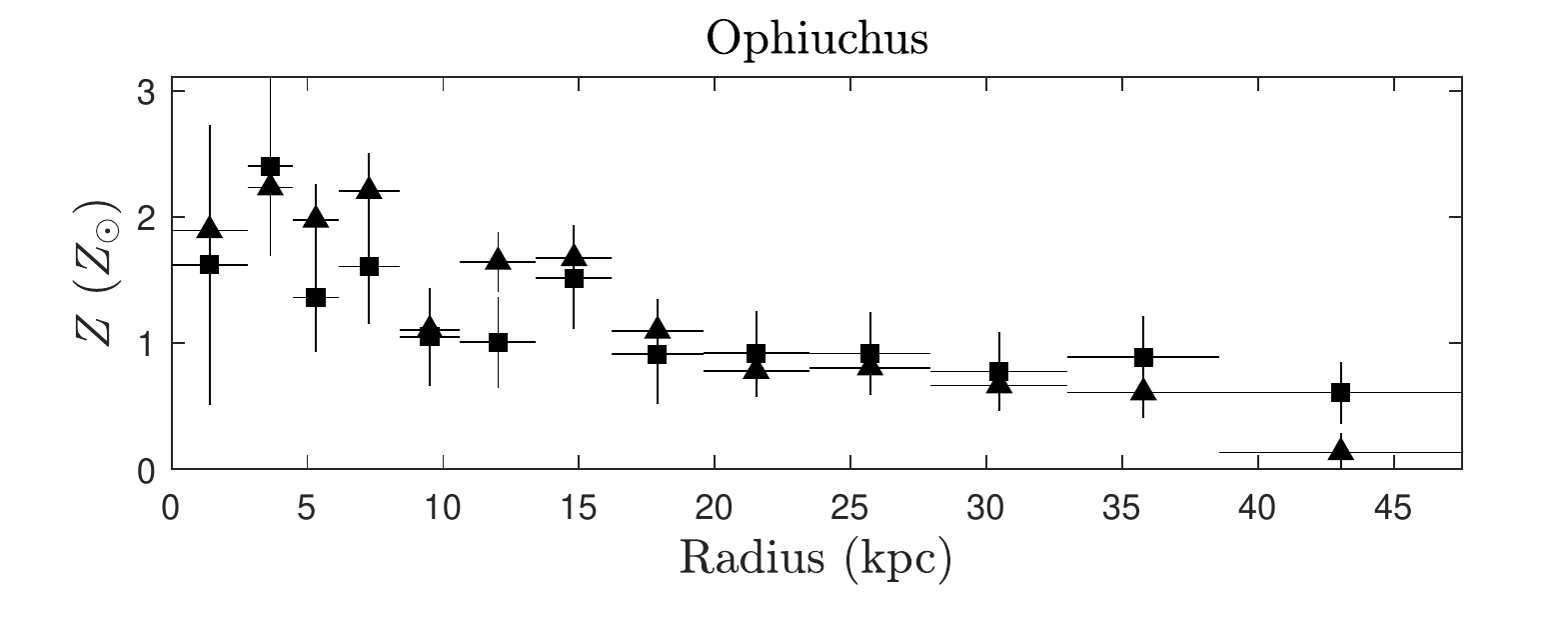}
\includegraphics[width=3.4in, height=1.45in, trim=80 410 80 200, clip]{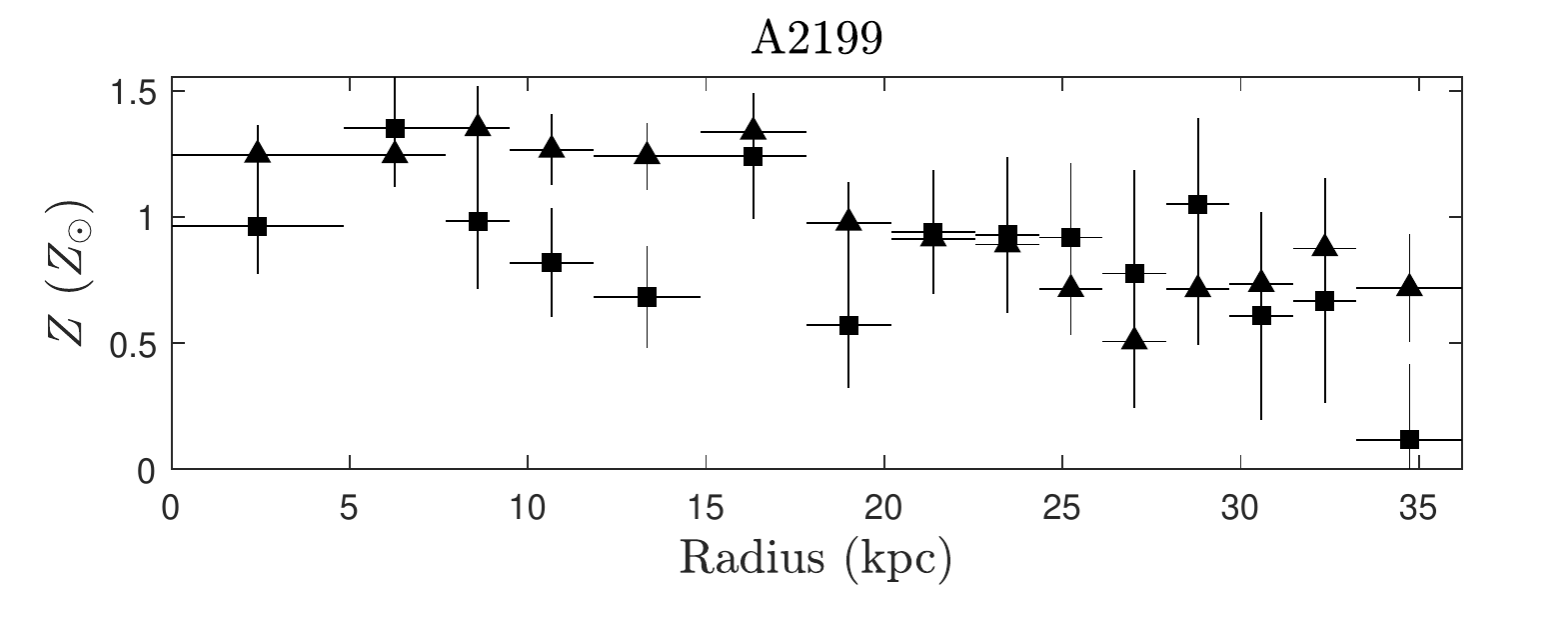}
\includegraphics[width=3.4in, height=1.45in, trim=80 410 80 200, clip]{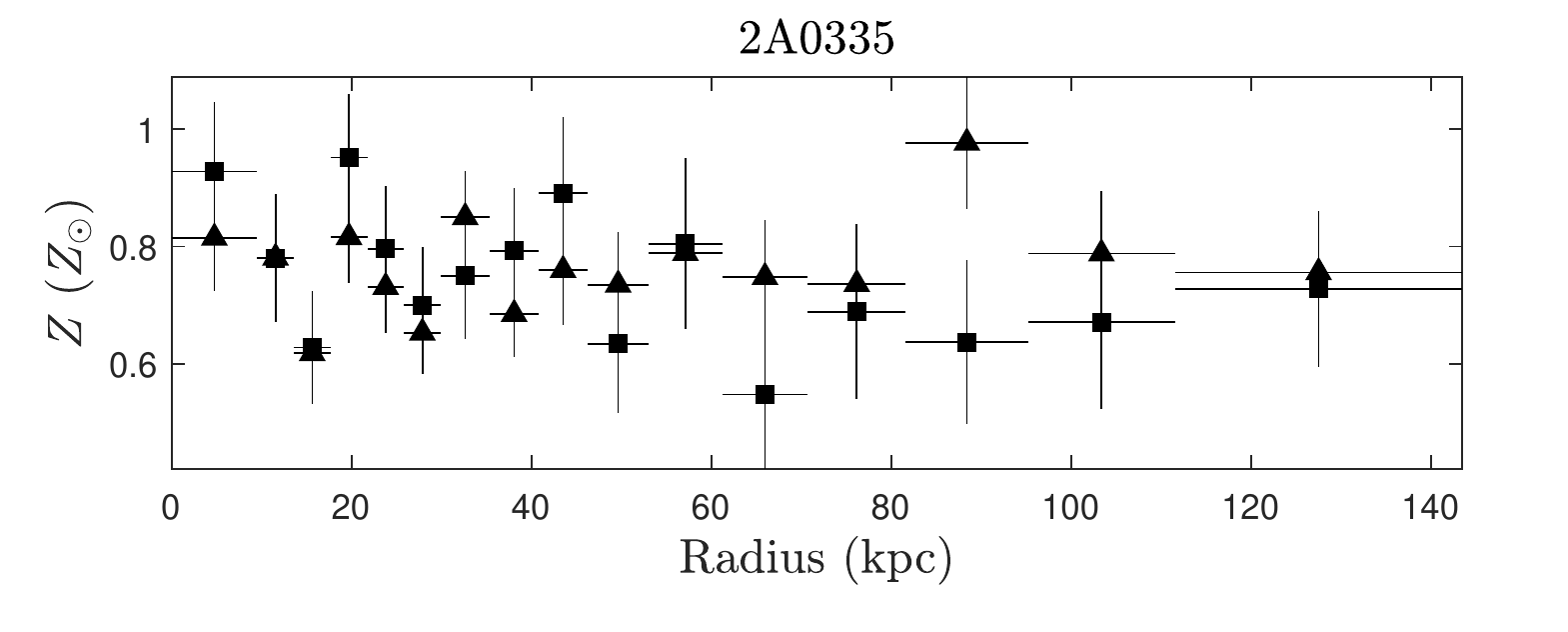}
\includegraphics[width=3.4in, height=1.45in, trim=80 410 80 200, clip]{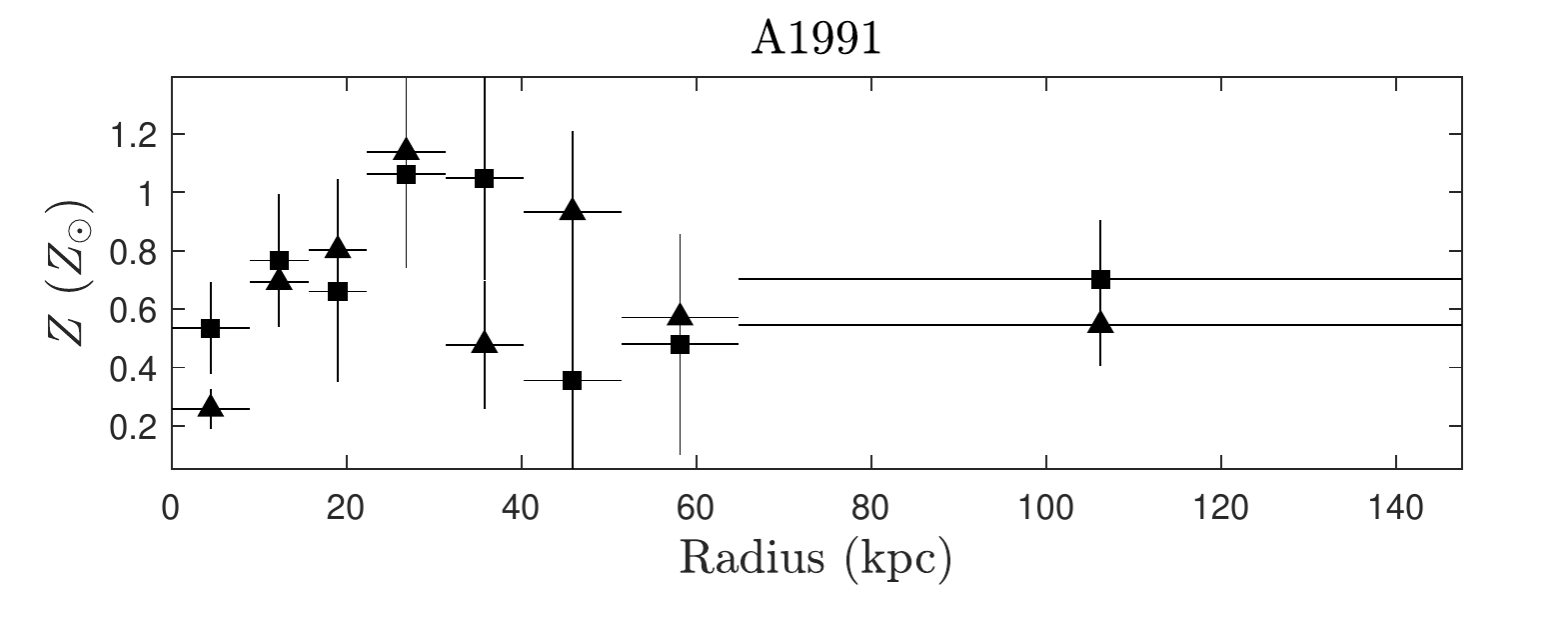}
\includegraphics[width=3.4in, height=1.45in, trim=80 410 80 200, clip]{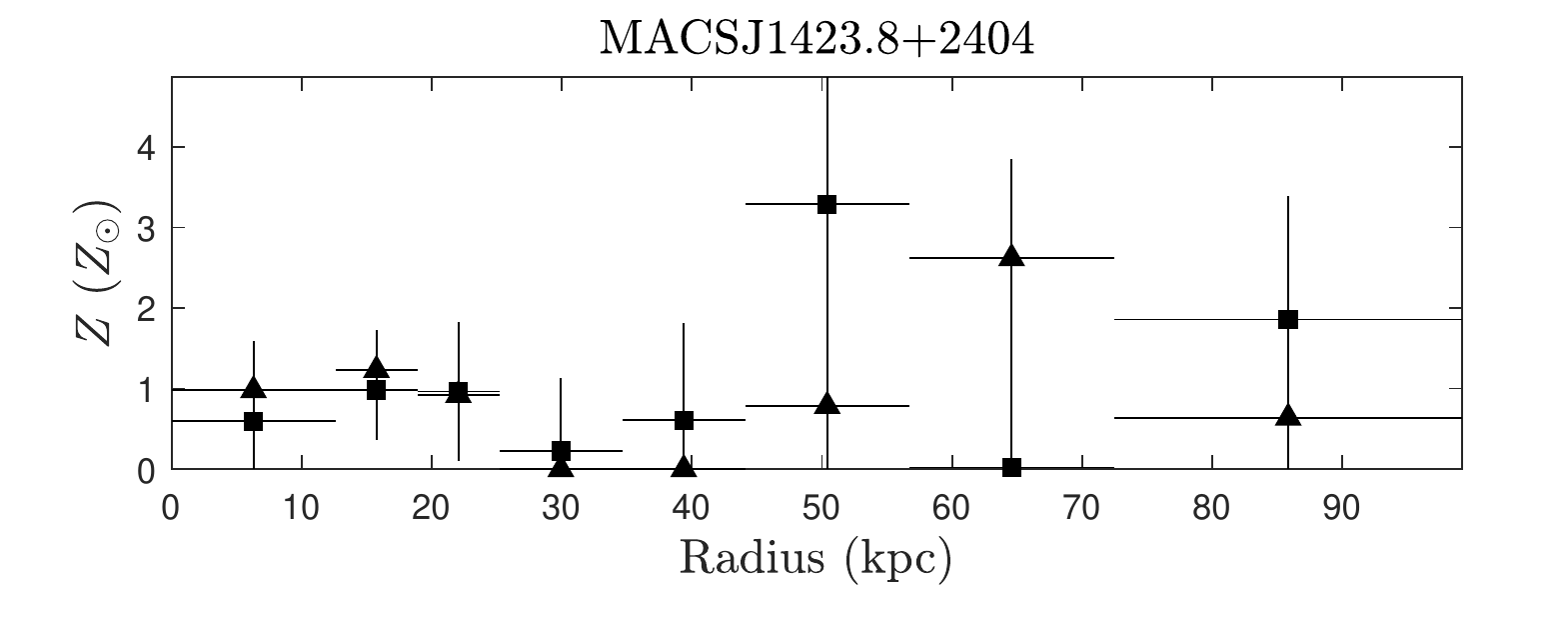}
\caption{Abundance profiles of Si (solid triangles) and S (solid squares) for all the groups and clusters
in our sample as a function of the physical radius in kpc. Abundances are expressed
in units of solar values as in \citet{asplund2009}.
}
\label{si_s}
\end{center}
\end{figure*}

The azimuthally-averaged, projected abundance profiles  of $Z_{\rm Fe}$, $Z_{\rm Ar}$ and $Z_{\rm Ne}$
for the 12 clusters in our sample are shown in Figure \ref{profiles}, while the profiles of Si
and S are shown in Figure \ref{si_s}. From a visual inspection, we see that we are able to confirm a
statistically significant Fe drop in most of the clusters, and to identify drop in the Ar abundance in
some of them. We also notice that the shape of the Ne profiles often appears to be consistent or at least
comparable with that of Fe and Ar, suggesting that Ne profiles are plausible, and that, despite
our measured Ne profile may be affected by unknown statistical uncertainties, we definitely have a
statistically-significant global detection of Ne across the sample in the core region.

Since the Fe profiles of these clusters have been already published, despite
with different analysis techniques and, in some cases, different data sets, we carefully check for
differences between our results and the literature.  If we compare our projected profiles with the
deprojected Fe profiles in \citet{panagoulia2015}, we find a good agreement, except in a few remarkable
cases. In NGC5813 we find a clear smooth drop in the innermost 5 kpc, opposed to the sharp drop within 5 kpc
in \citet{panagoulia2015}.  We also find a smoother and more pronounced iron drop in A3581, which is
considered only a possible drop in \citet{panagoulia2015}, despite their large error bars create no tension
with our profile. While these differences may be due to the deprojection procedure, the results for 2A0335
are dramatically discrepant.  In this cluster we find an iron peak opposed to the
well defined iron drop shown in \citet{panagoulia2015}.  We actually find that switching from 1T {\tt apec} model
to our reference 2T {\tt apec} model is critical to find the peak.  We also find a strong dependence of the
temperature profile on the $NH_{Gal}$ value, which we measure to be $3.1\times 10^{21}$ cm$^{-2}$ as in
\citet{pinto2015}, as opposed to lower value used for this clusters in \citet{panagoulia2015}. The value
of $NH_{Gal}$ is found to strongly affect the Fe abundance particularly for a 2T {\tt apec} model.
Our result is in agreement with \citet{werner2006}.  However, the peculiar nature of the core of 2A0335,
characterized by bright blobs and a complex metallicity map in the 2D analysis \citep[see][]{2009Sanders}
suggests that assuming azymuthally average value for the ICM may provide contradictory results
depending on details of the analysis, including the choice of the center (we assumed the BCG position in
this case). We conclude that 2A0335 should not be included among the list of clusters showing a central
Fe drop.  We also find consistency with \citet{lakhchaura2018} for the Fe and Ar profiles of the Centaurus
cluster, despite the different choice of the size of the radial bins hampers a detailed comparison.
We also find that our projected Fe profile of the Ophiuchus cluster and the deprojected one
in \citet{million2010} are compatible.  Despite the different angular resolution makes it very difficult the
comparison of {\sl Chandra} and XMM-Newton profiles in the innermost regions, we find a broad
consistency with the Fe profiles of NGC4636, NGC5044, NGC5813 measured with XMM-Newton in \citet{mernier2017},
while we notice some differences in the cases of A3581, NGC5846 and Perseus.  Significant difference, instead,
are found for A262 and A1991, which were, in fact, not classified as Fe-drop by \citet{mernier2017}.
Overall, we find that the Fe profiles obtained with our approach are broadly consistent with literature,
while significant differences occur in particular cases with extremely complex core structure, or in data
with lower angular resolution.

\begin{figure*}
\includegraphics[width=3.4in, height=3in, trim=80 180 80 200, clip]{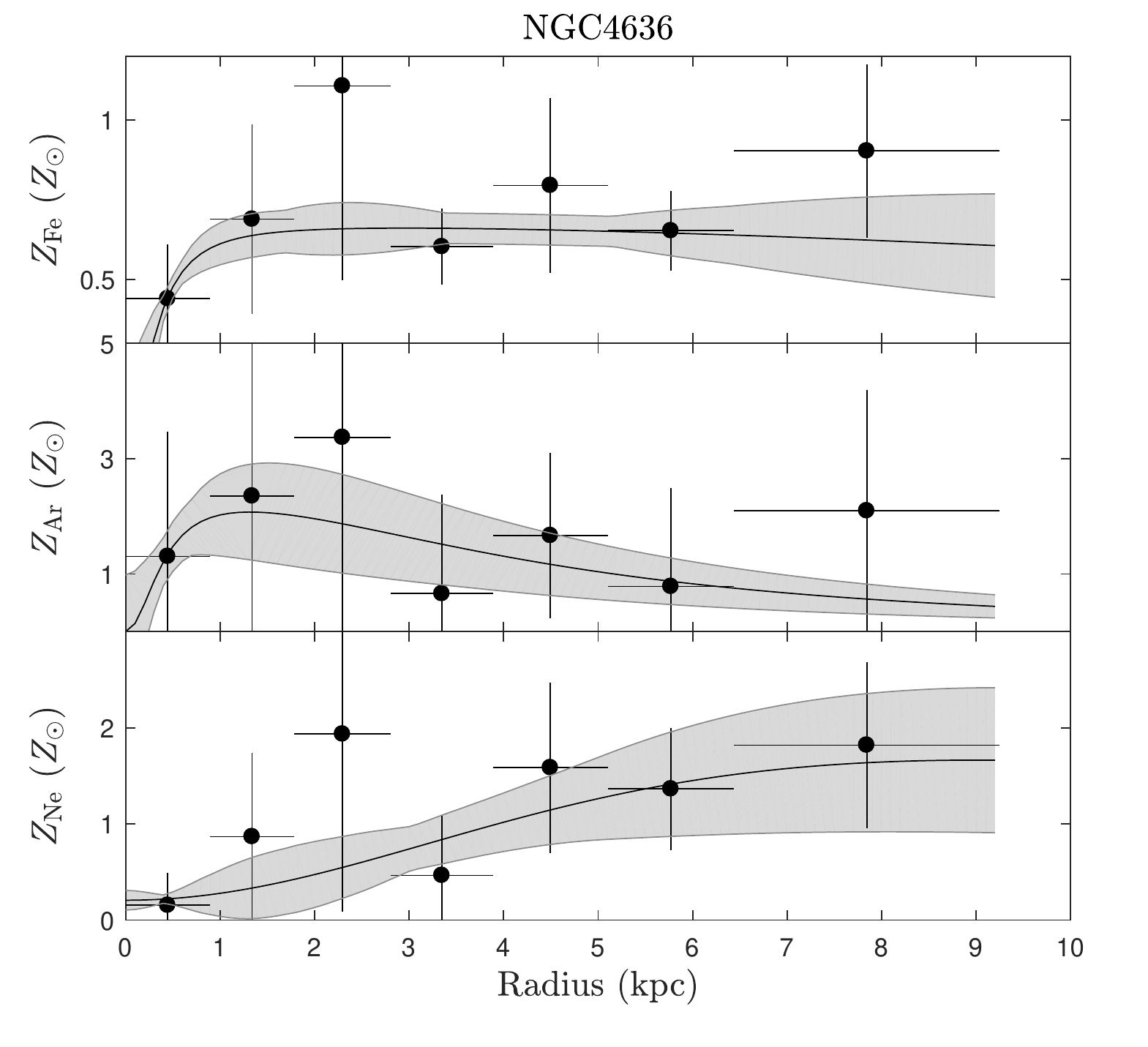}
\includegraphics[width=3.4in, height=3in, trim=80 180 80 200, clip]{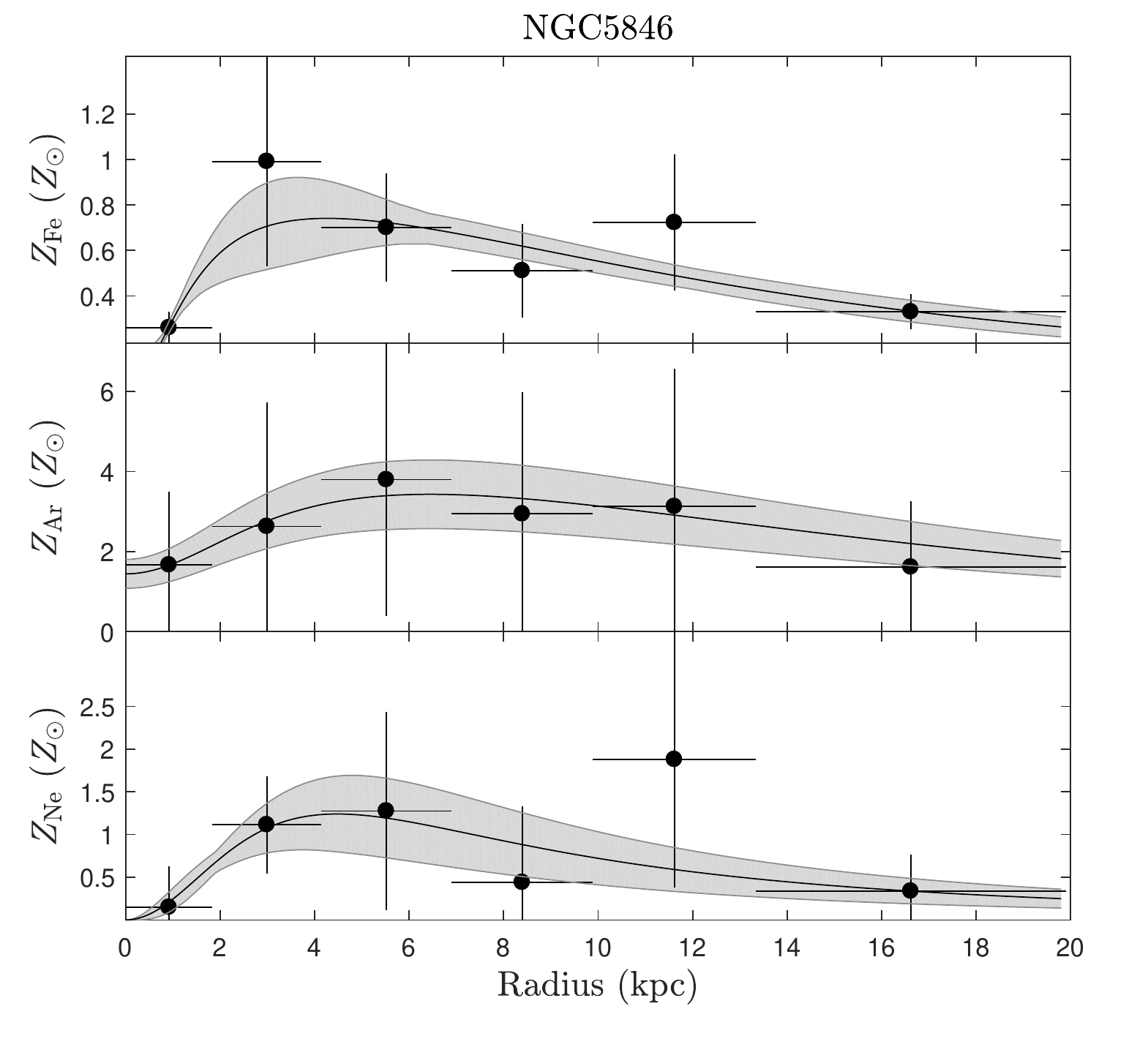}
\includegraphics[width=3.4in, height=3in, trim=80 180 80 200, clip]{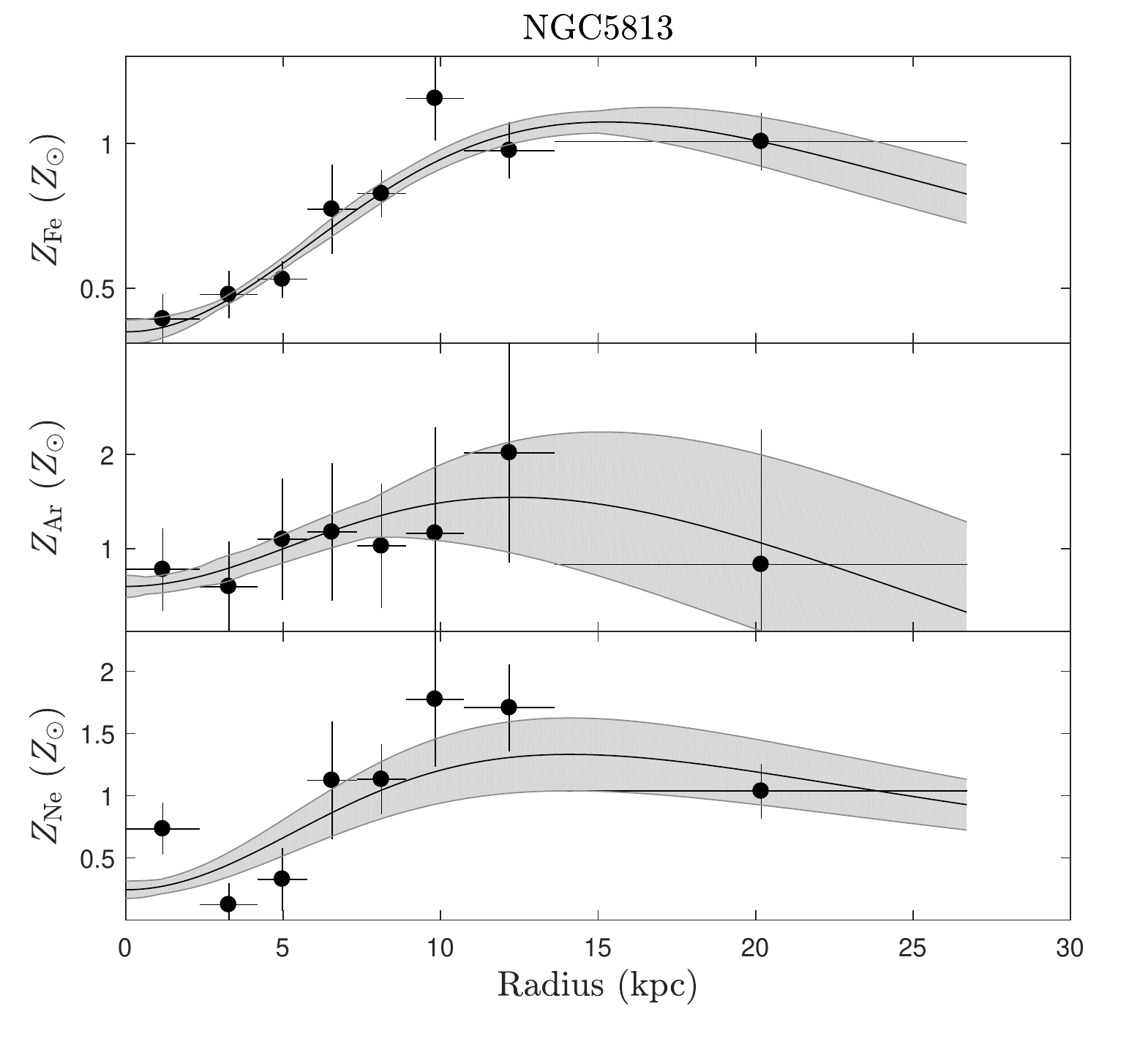}
\includegraphics[width=3.4in, height=3in, trim=80 180 80 200, clip]{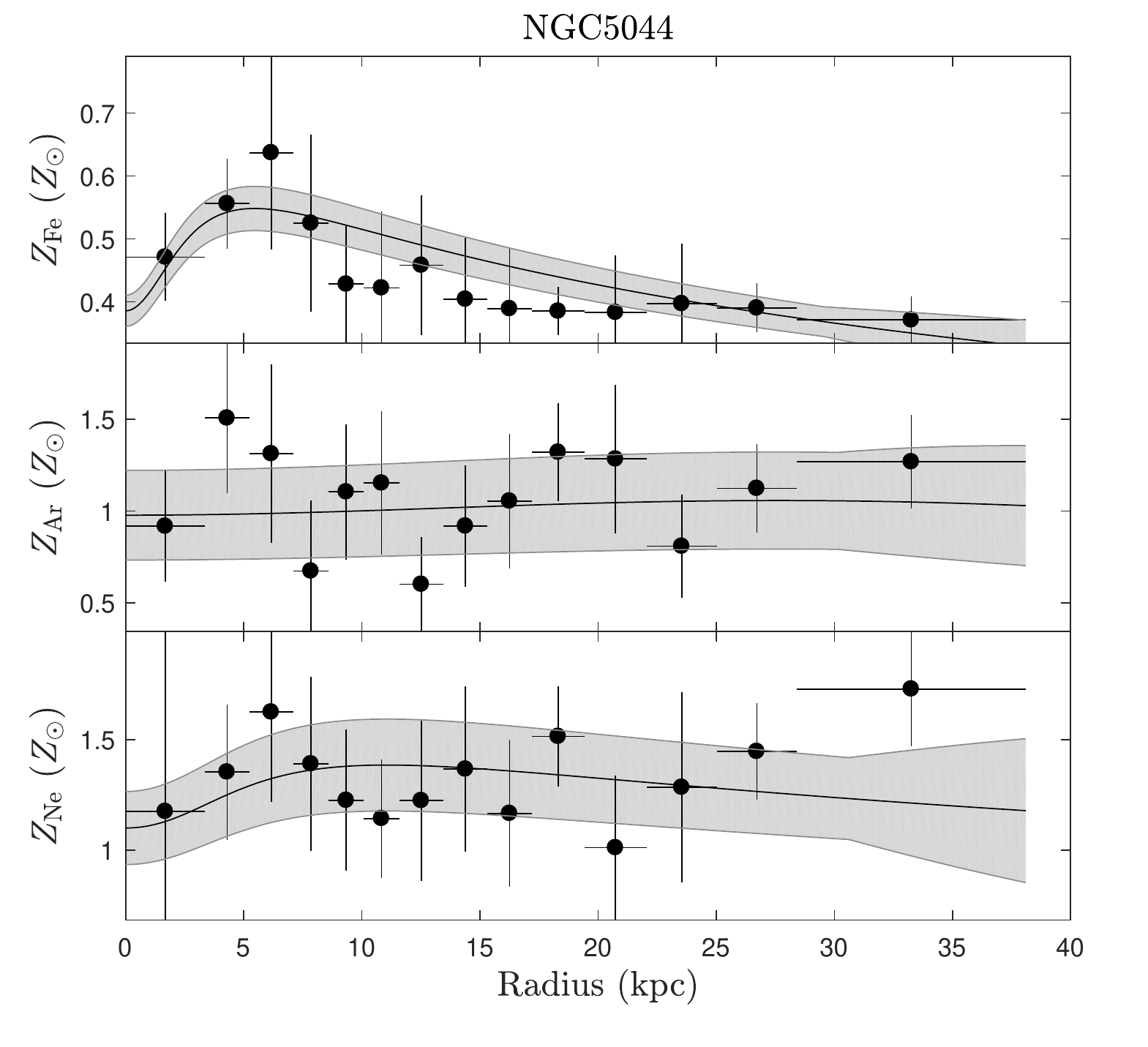}
\includegraphics[width=3.4in, height=3in, trim=80 180 80 200, clip]{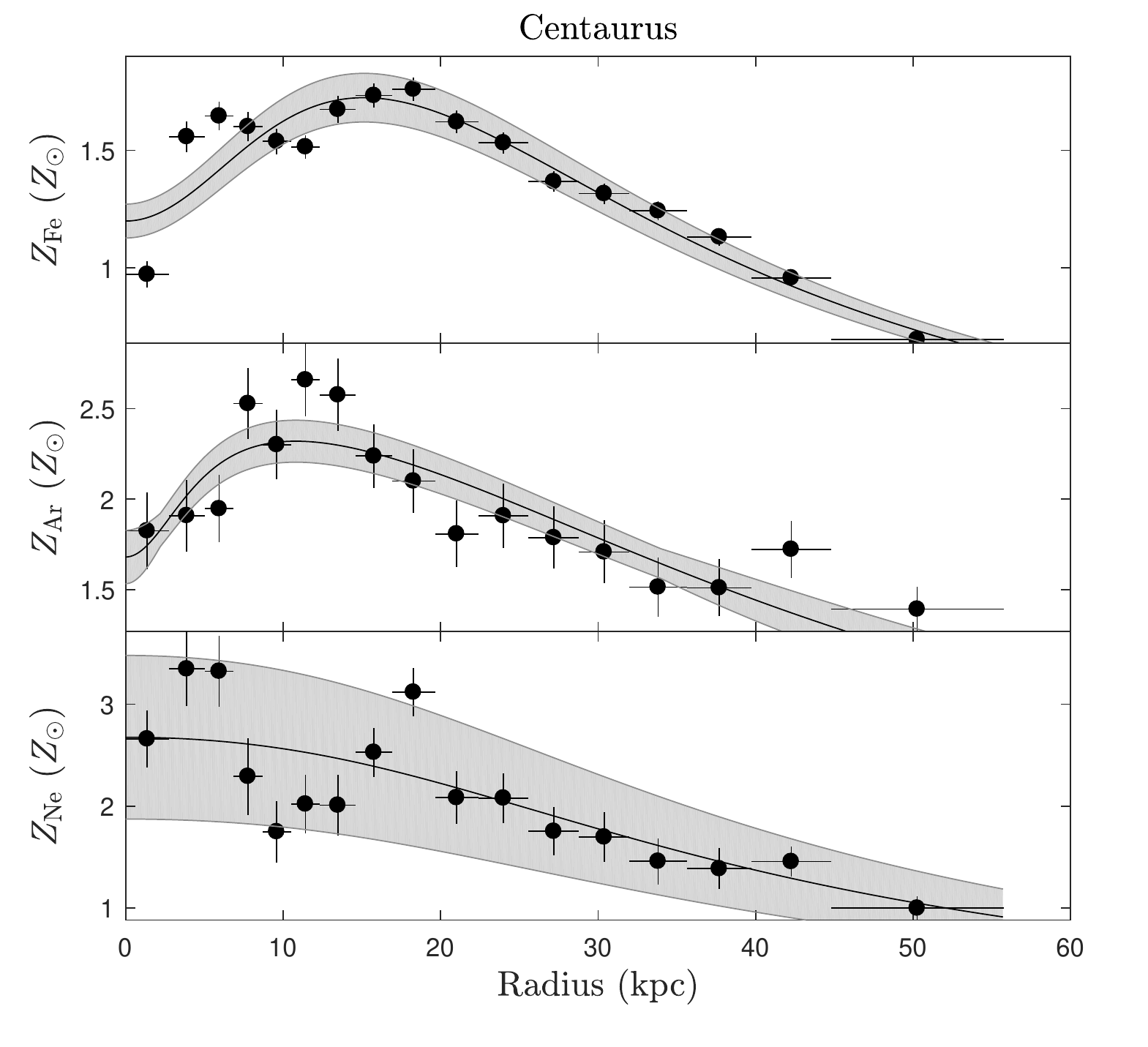}
\includegraphics[width=3.4in, height=3in, trim=80 180 80 200, clip]{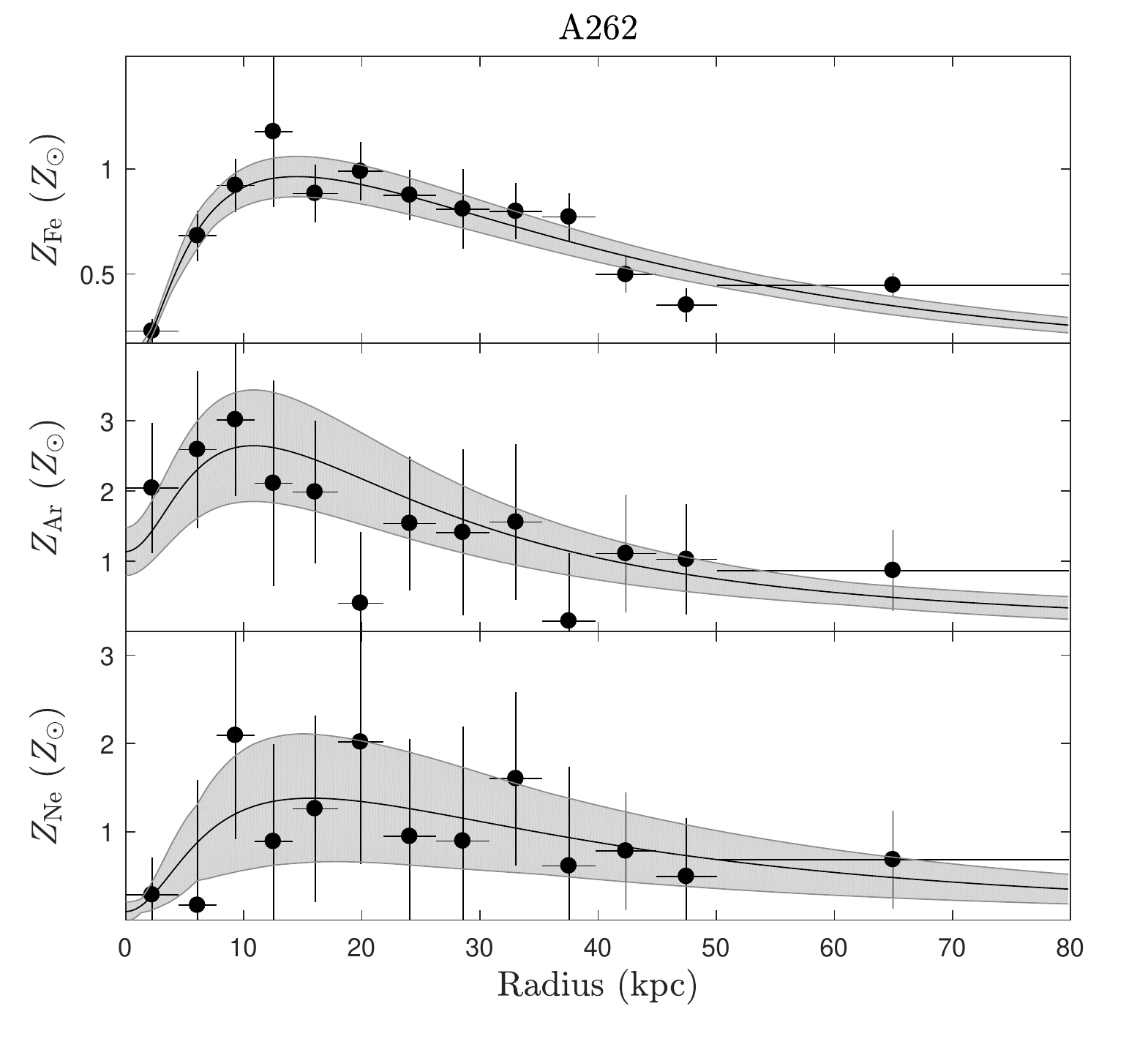}
\caption{The abundance profiles of Fe (top), Ar (middle), and Ne (bottom) for all the groups and clusters
in our sample, as a function of the physical radius in kpc. Abundances are expressed
in units of solar values as in \citet{asplund2009}. Solid line and shaded area show the best fit
function given in equation \ref{eq1}, and the 1$\sigma$ confidence interval, respectively.}
\label{profiles}
\end{figure*}

\begin{figure*}
\contcaption{}
\includegraphics[width=3.4in, height=3in, trim=80 180 80 200, clip]{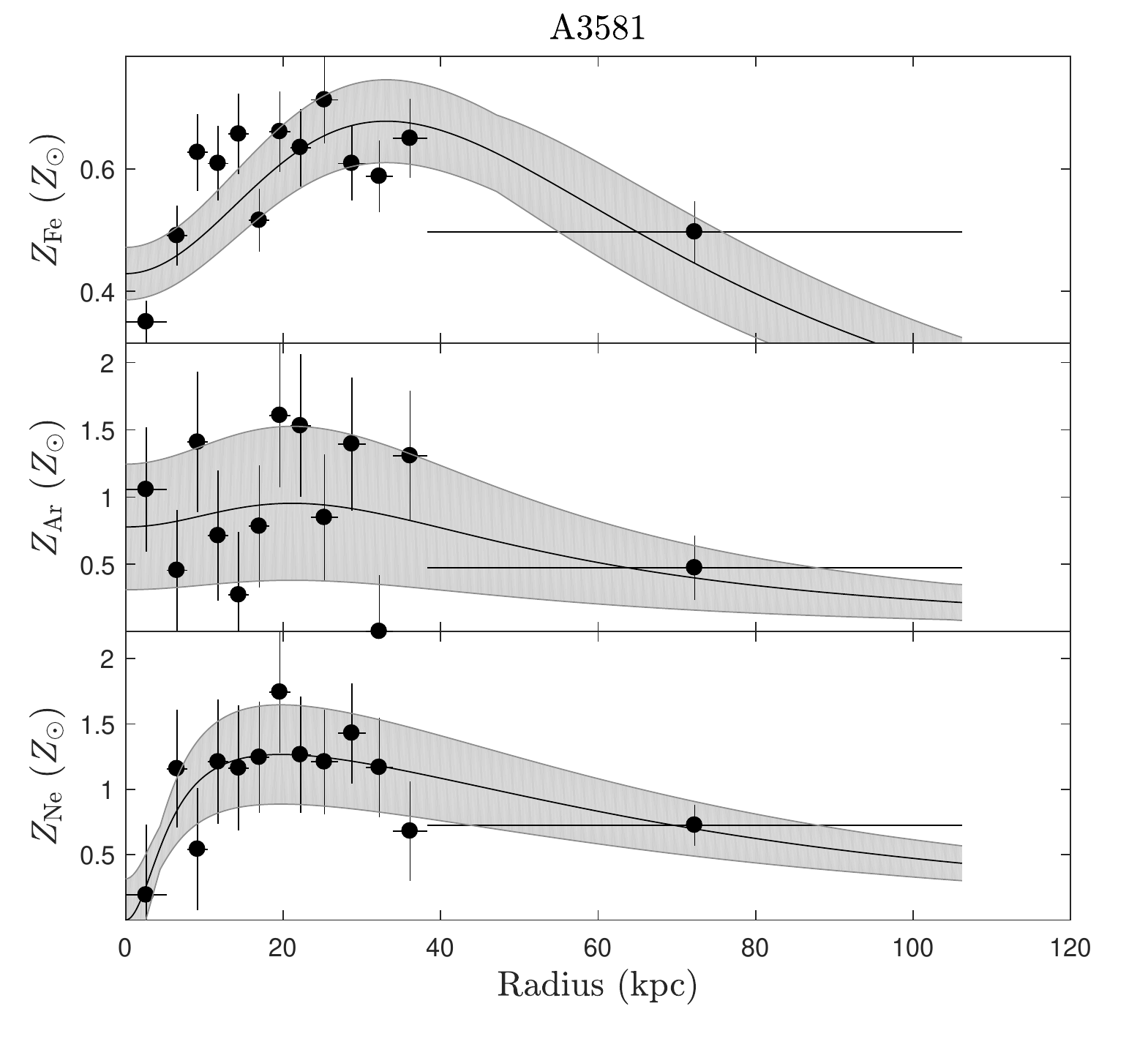}
\includegraphics[width=3.4in, height=3in, trim=80 180 80 200, clip]{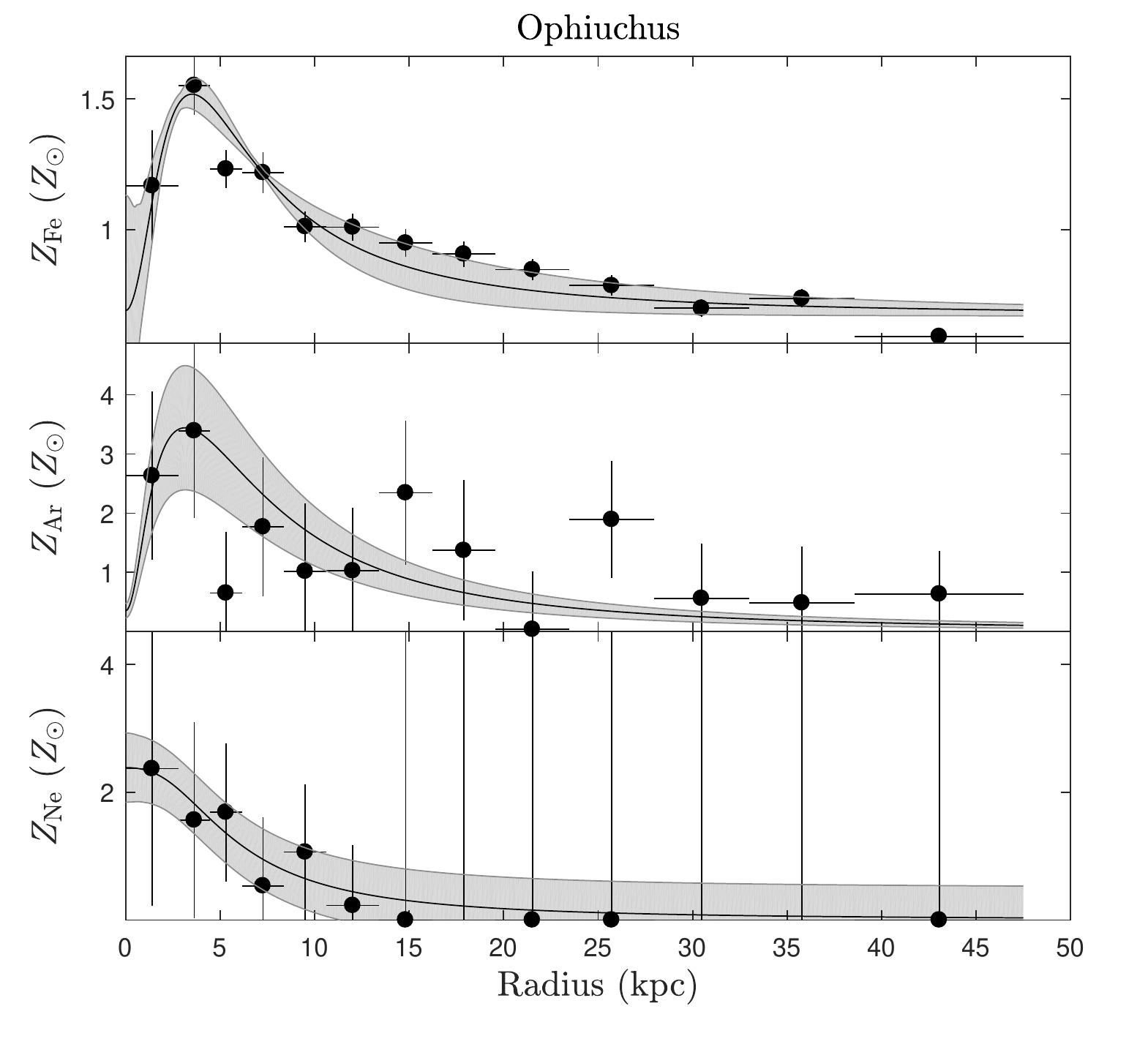}
\includegraphics[width=3.4in, height=3in, trim=80 180 80 200, clip]{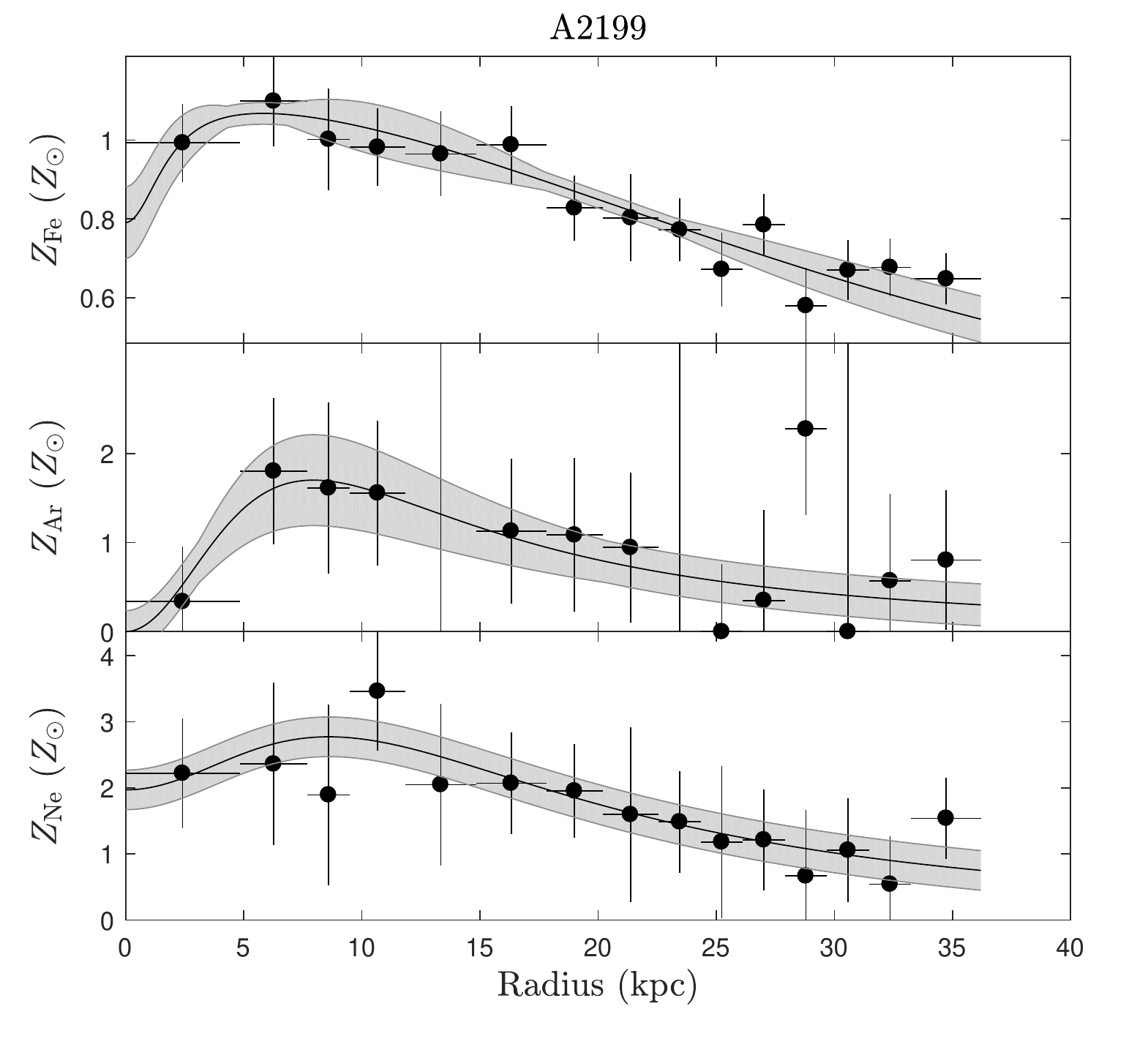}
\includegraphics[width=3.4in, height=3in, trim=80 180 80 200, clip]{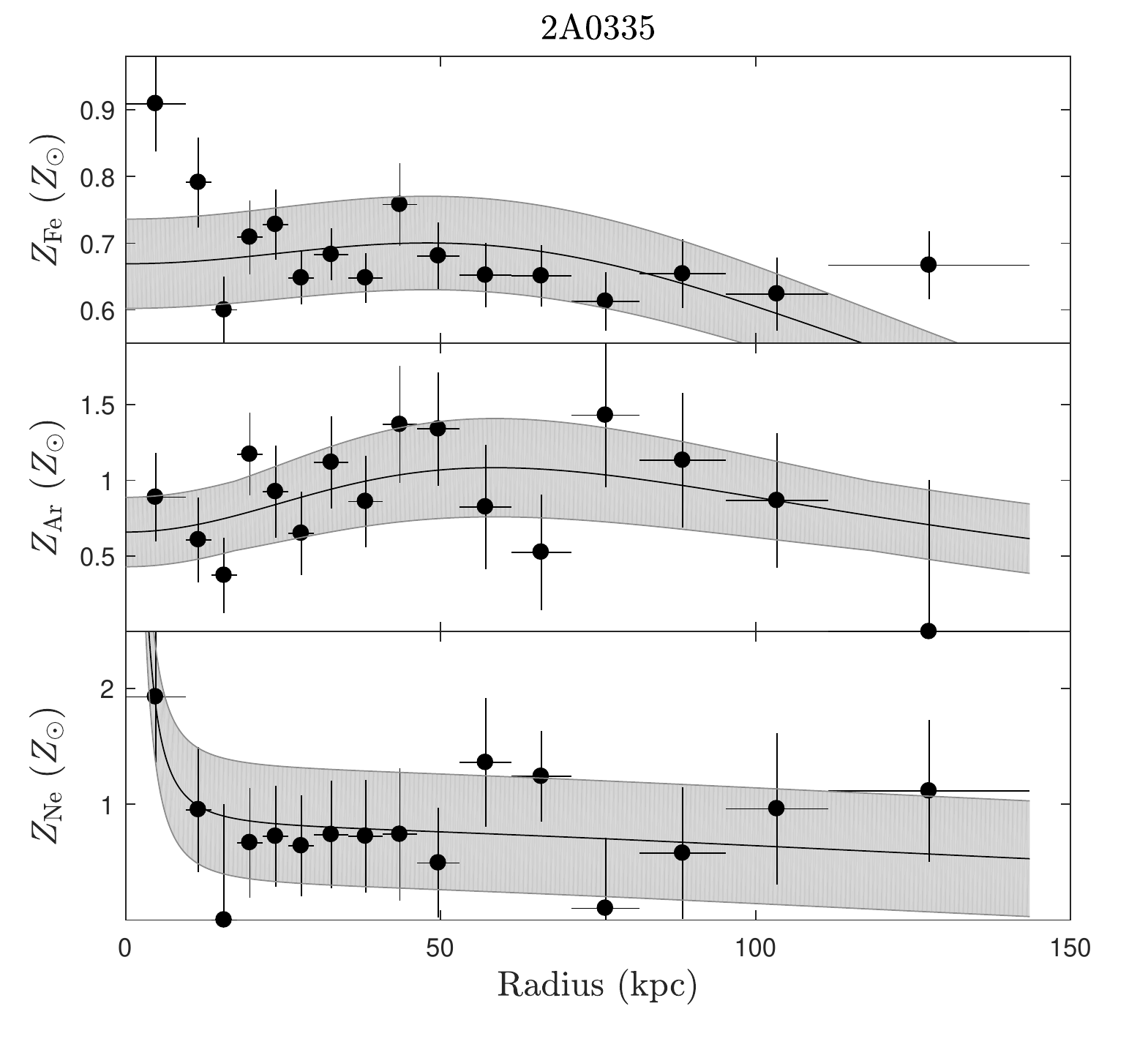}
\includegraphics[width=3.4in, height=3in, trim=80 180 80 200, clip]{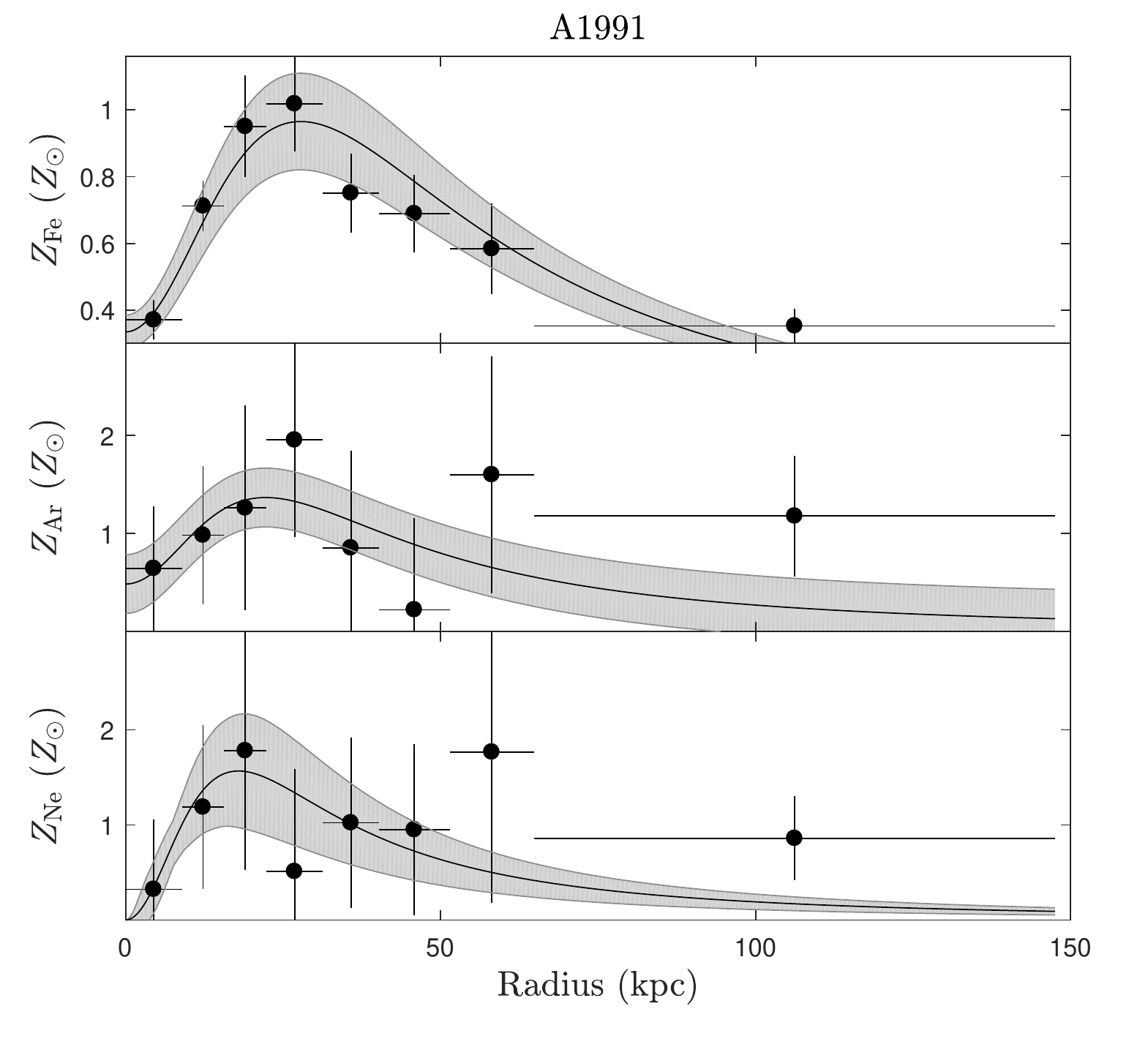}
\includegraphics[width=3.4in, height=3in, trim=80 180 80 200, clip]{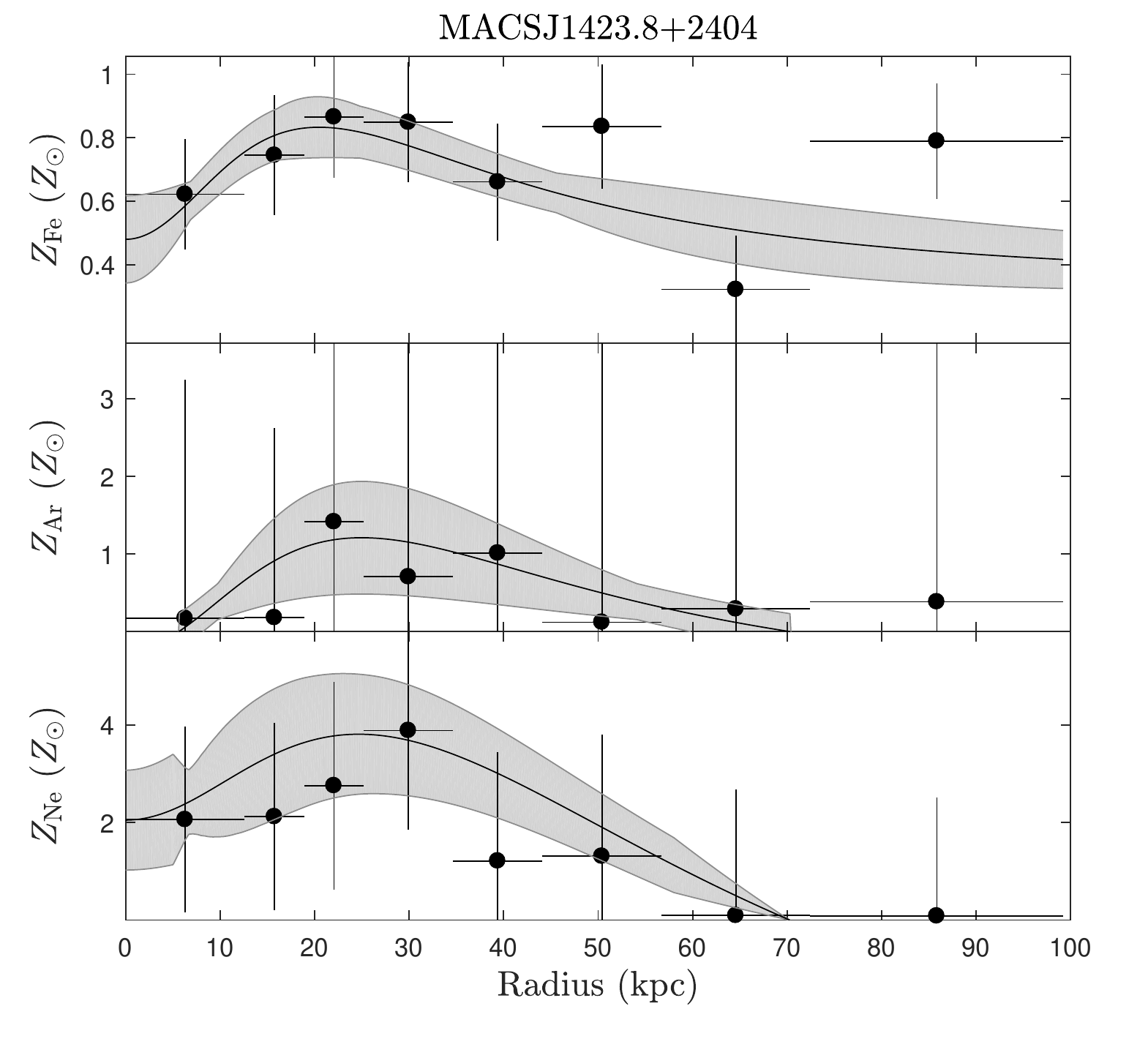}
\end{figure*}

\begin{figure}
\begin{center}
\includegraphics[width=3.4in, height=3in, trim=95 210 100 230, clip]{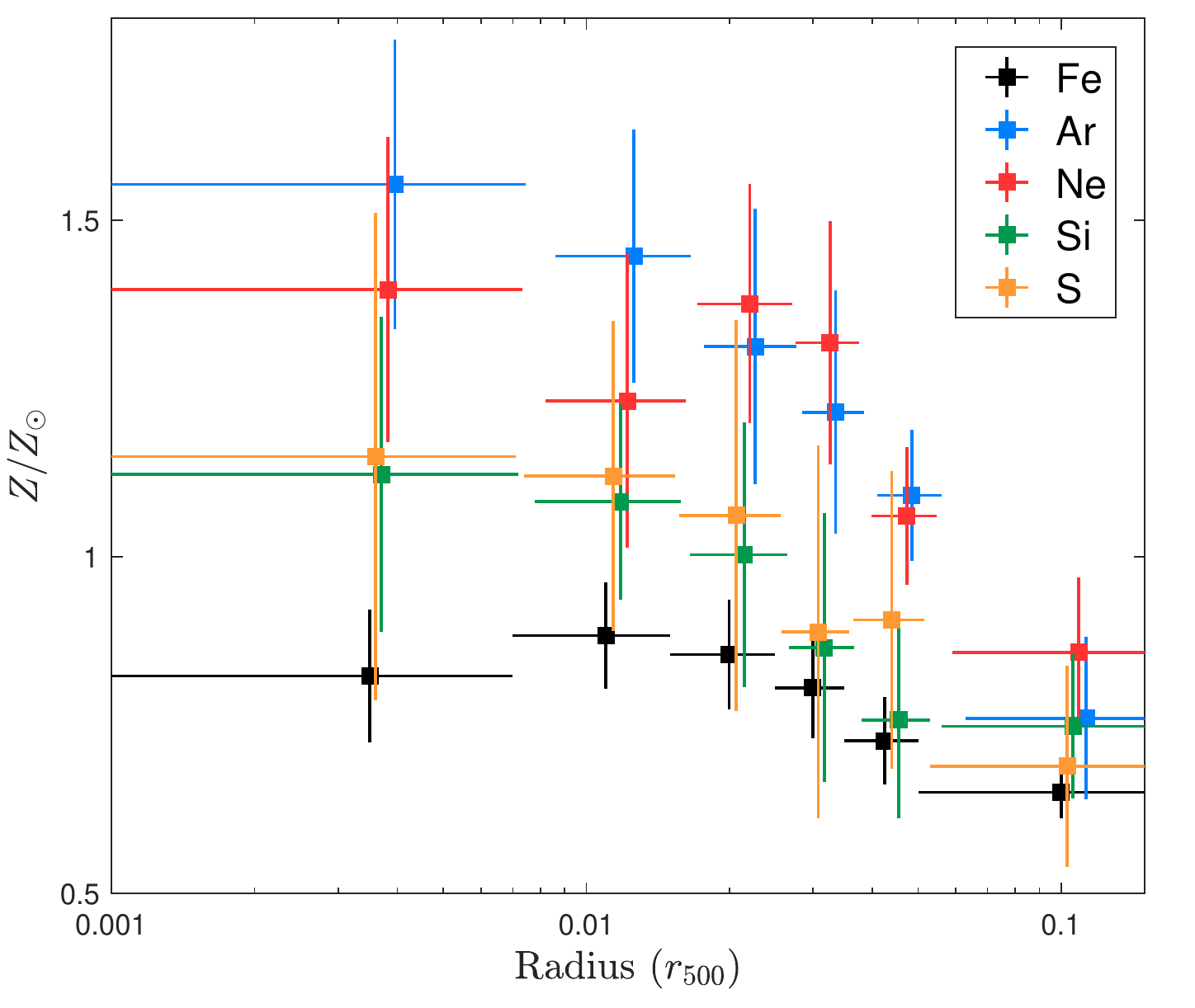}
\includegraphics[width=3.4in, height=3in, trim=95 210 100 230, clip]{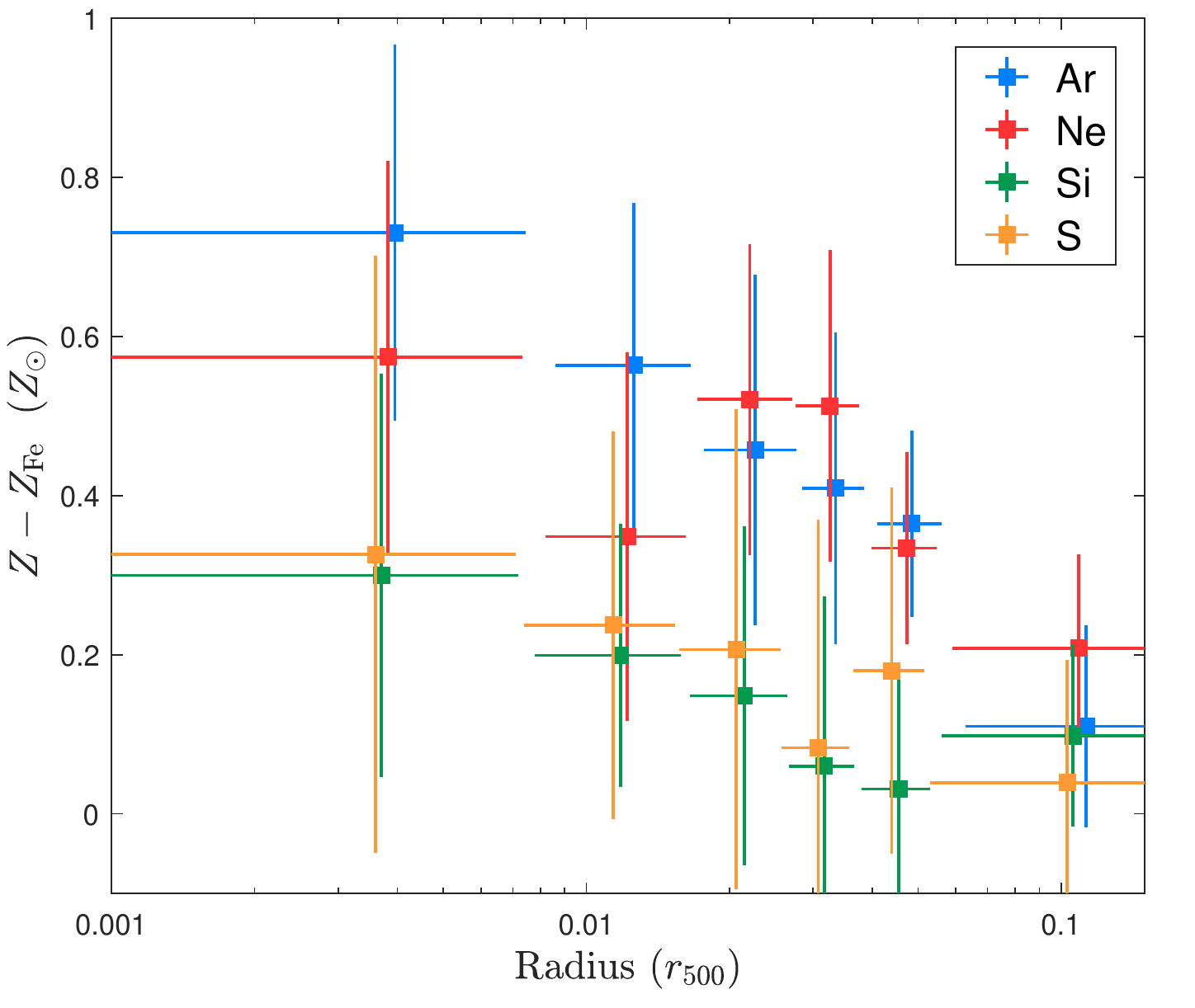}
\caption{Upper panel: the stacked abundance profiles of Fe, Ar, Ne, Si, and S in our sample,
as a function of the rescaled radius $r/r_{500}$.
Bottom panel: abundance excess profiles of Ar, Ne, Si, and S relative to Fe. The data points have been slightly shifted on the x-axis for clarity.
}
\label{all}
\end{center}
\end{figure}

We note, however, that the precise identification of the central drop is not straightforward and may be rather
uncertain if based on, e.g., the distance between the highest and the lowest values in each profile, since
it would be sensitive to the width of the annuli, and blurred by the statistical noise. Therefore, to better
quantify the abundance drops in a uniform way, we fit each profile with an empirical function with five
free parameters, able to reproduce a smooth, non-monotonic profile:
\begin{equation}
Z(r) = A\cdot\frac{\frac{r}{B}+C}{\frac{r}{B}+1}\cdot\frac{1}{(1+(\frac{r}{D})^2)^E}\, ,
\label{eq1}
\end{equation}

\noindent
where the parameters $B$ and $D$ are the spatial scales corresponding to the drop in the inner few tens
of kpc, and the outer profile, respectively, while the parameter $C$ corresponds to the minimum value
in the center.  This functional shape is a simplified version of the profile used to described the temperature
in \citet{vikhlinin2006a}, and it is overdetermined given our data quality.  However, this function
can reasonably reproduce the behaviour of all the different profiles.
We stress that the parameters in the function have no direct physical meanings, but allow us to obtain a
fit to all the profiles and the associated uncertainty as a function of the physical radius.
The fit of the profile and the uncertainty on $\Delta Z$ are obtained with a Monte Carlo approach.
Eventually, we visually inspect the profiles to search for any poor fitting and
check the $\chi^2$ values, finding that in all the cases we obtain acceptable fits. We remark that
in the fits we allowed the minimum abundance values in each bin to be negative, an aspect that hardly
affect single-cluster profiles (only a few points in Ophiuchus and A2199 turn out to be negative)
but is relevant particularly in the stacking procedure, that may be slightly biased high if all the values
are forced to be positive.

From the the best-fit analytic function, we can directly quantify the central abundance drop as
$\Delta Z = Z_{\rm peak}-Z_{\rm in}$, where $Z_{\rm in}$ is the fitted abundance at the radius of the
innermost bin, and $Z_{\rm peak}$ is the maximum value of the fitted abundance.
The uncertainty on $\Delta Z$ is obtained by assuming the 1$\sigma$ uncertainties of $Z_{\rm peak}$ and $Z_{\rm in}$ and summing them in quadrature.
A central drop is identified only when $\Delta Z$ has a significance level higher than 2$\sigma$.
The measured values of $\Delta Z$ are shown for Fe, Ar and Ne in the second, third and fourth
column, respectively, of Table \ref{result}.  Our results confirm that all the clusters in our sample exhibit
a central Fe abundance drop at more than 2$\sigma$ c.l., except for A2199 and 2A0335.
However, we find a significant Ar drop in only 4 sources (NGC5813, Centaurus, A2199, and A1991) at more
than 2$\sigma$, despite we have hints of a drop in most of the other sources.  Formally,
we also find a central drop in Ne abundance in 4 clusters (NGC5846, NGC5813, A3581, and
A1991) at more than 2$\sigma$. We notice that, given the larger uncertainties in the profiles
of Ar and Ne, our results are consistent with an abundance drop common to the three elements,
which suggests a mechanical process removing the highly enriched ICM from the innermost region as an explanation
for the observed profiles.  We also remark that the detection of abundance drop in
MACSJ1423.8+2404 at $z=0.543$ implies that this feature is also present in high redshift clusters,
and that the mechanism generating it has already taken place at a look-back time of
$\sim$6~Gyr.

We can push further our analysis by stacking the abundance profiles of each element after
rescaling the radius of each cluster by $r_{500}$.  We divide the [0--0.15]~$r_{500}$ interval into 6 bins
choosing the width in order to have a comparable number of data points in each bin. The values of $r_{500}$ are
taken from the compilation of clusters presented in \citet{pinto2015} and \citet{liu2018}.
We then average all the data points in each bin weighted by the inverse squared errors and by the overlap of the
geometric area of each physical annulus with the final bin, following the method of \citet{leccardi2008}.
In addition, we also allow the free
abundance parameters to be negative for the stacking, since this is the only way to properly deal with noise. As discussed in \citet{leccardi2008}, forcing the abundance to be always positive may result in a positive statistical bias when averaging out the profiles.
In the investigation of the stacked profiles we consider also S and Si, whose single-cluster profiles
are shown in Figure \ref{si_s}.

The stacked abundance profiles of Fe, Ar, Ne, Si, and S are shown in the
top panel of Figure \ref{all}. We find that in the central regions, the average abundances of Ar and Ne in
our sample are significantly higher than that of Fe, while they tend to be consistent in the outermost
bin, corresponding to [0.05--0.15]~$r_{500}$. On the other hand, the profiles of Si and S are consistent
with Fe within $1\sigma$, despite they tend to show a slight increase. However, the most interesting
and convincing feature is the much larger gradient of the Ar and Ne profiles
with respect to Fe, Si and S, which are all elements that may be depleted into cold dust grains.
It is useful to remind that at larger radii, Ar abundance has been shown to be comparable or lower with
respect to Fe \citep[see][]{mernier2017}. The same behaviour can be seen in the absolute difference
of the abundance of different elements as a function of radius, which is shown in the lower panel of
Figure \ref{all}. We find that the difference $Z_{\rm Ar}-Z_{\rm Fe}$ decreases systematically with
radius, while $Z_{\rm Si}-Z_{\rm Fe}$ and $Z_{\rm S}-Z_{\rm Fe}$ show a much milder decrease with radius.
We cannot draw any conclusion on the quantity  $Z_{\rm Ne}-Z_{\rm Fe}$ given the large uncertainties,
not to mention the unknown systematics. Finally, we verified {\sl a-posteriori} that the stacked profiles
are not dominated by the Centaurus cluster, despite it has by far the highest quality spectra.  The stacked
profiles obtained after excluding the Centaurus cluster have only negligible differences with respect to
the results shown in Figure \ref{all}.

\begin{figure}
\begin{center}
\includegraphics[width=3.4in, height=3in, trim=85 210 95 225, clip]{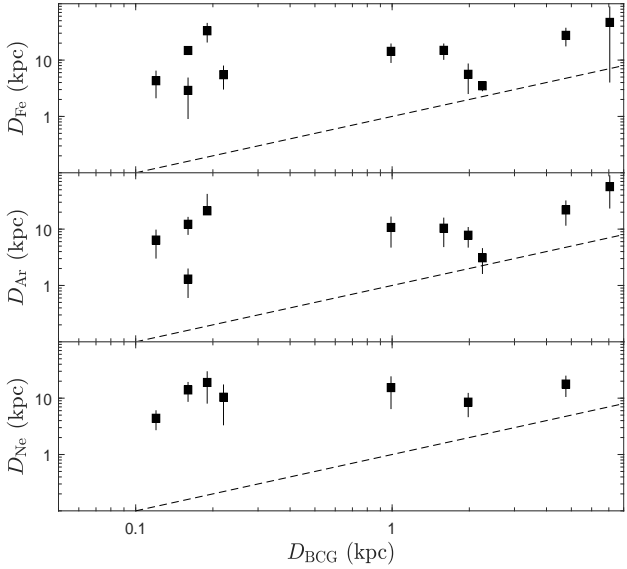}
\caption{The displacement of the abundance peak with respect to the X-ray centroid are shown for
Fe, Ar and Ne as a function of the offset between the BCG position and the X-ray centroid $D_{\rm BCG}$.
The dashed line shows the $D_Z-D_{\rm BCG}$ relation as a reference.}
\label{bcg}
\end{center}
\end{figure}

In Table \ref{result} we also list the distance of the abundance peak of each element from the peak in the
X-ray surface brightness, and, in the last column, the displacement of the BCG from the X-ray peak.
We unambiguously identify the BCG of all the clusters but
MACSJ1423.8+2404. The coordinates of the BCG come from the 2MASS catalog or NED/SIMBAD. The position of
the abundance peak is measured from the best-fit of the abundance profile of the corresponding element.
For most of the clusters, we find that $D_{\rm BCG}$ is much smaller than the distance from the abundance peak
to the cluster center, implying that the BCG is inside the `hole' carved into the abundance distribution.
There is only one exception: the Ophiuchus cluster, with $D_{\rm BCG}$=2.25~kpc, and
$D_{\rm Fe}$=$3.5\pm1.1$~kpc. This is consistent with the clear signatures of an ongoing merger
found by \citet{million2010}. For this cluster, we recompute the abundance profiles after
setting the center on the position of the BCG, and we obtain a consistent result.
In Figure \ref{bcg} we show the relation between $D_{\rm Fe}$, $D_{\rm Ar}$, $D_{\rm Ne}$ and $D_{\rm BCG}$.
We find no statistically significant trend among the BCG displacement and the position of the peak of
the abundance profiles. Therefore, we conclude that there is no association between the displacement of
the abundance peak (as a proxy for the size of the abundance drop region) and the offset of the BCG with respect to the X-ray centroid in our sample.

To summarize, we find 4 clusters with a drop in the Ar abundance profile at more than 2$\sigma$ c.l.,
compared to 10 clusters with an Fe drop at the same significance level. This indication of a different
spatial distribution
among this two elements is further strengthened by our analysis of the stacked profiles, where the normalization
and the slope of the Ar profile are clearly different from that of Fe, implying a larger Ar abundance in the
core regions at radii $\leq 0.05~r_{500}$.  Possibly, the same trend is present in the Ne profile.
At the same time, we find overall consistency between the profile of Fe, Si, and S,
apart from a modest enhancement, confirming a significant difference in the spatial distribution of elements
that may be depleted into dust grains, and of noble elements.
Therefore, despite our analysis confirms that the abundance drop is a characteristic shared by all the elements,
suggesting that the uplifting of high metallicity gas driven by bubbles is an effective process,
we also found a significant difference between Ne and Ar and the other elements.  This result implies that
the depletion of Fe into dust grains during the baryon cycle in cool-core groups and clusters
is also an effective process in shaping the abundance profiles. Since here we focus on projected values,
further assumptions and a detailed modelization are both
needed before quantifying the effects of dust depletion in terms of Fe mass-loss, a task which goes beyond
the scope of this paper.

\section{Discussion: robustness of the spectral analysis}

\begin{figure}
\begin{center}
\includegraphics[width=3.4in, height=3in, trim=85 210 100 225, clip]{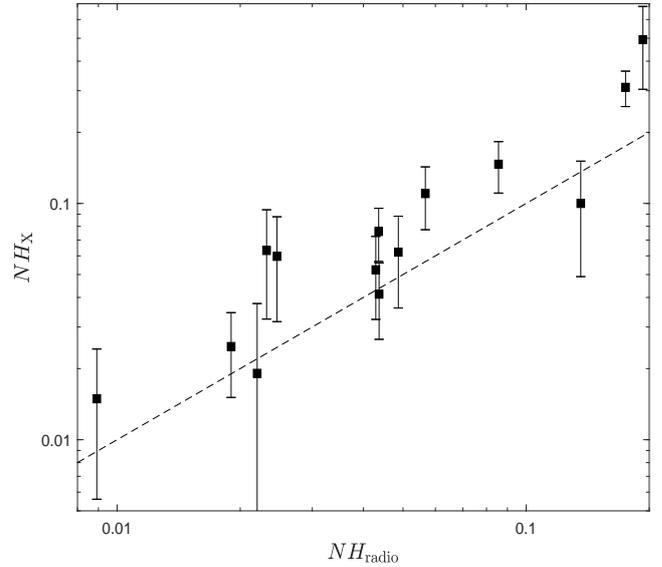}
\caption{The Galactic HI column density from \citet{2005Kalberla} compared to the average
$NH_{Gal}$ best-fit values obtained in each source of our sample. The dashed line shows the
$NH_{\rm X} = NH_{\rm radio}$ relation.}
\label{nh}
\end{center}
\end{figure}

In this section we discuss some aspects of our spectral analysis which may potentially introduce fitting bias.
The first aspect is the choice of leaving the Galactic absorption parameter $NH_{Gal}$ free to vary
in each bin.  It is a common choice in X-ray spectral fitting to freeze $NH_{Gal}$to the value of
\citet{2005Kalberla}.
However, given the high signal of our spectra, and the complex thermal structure in each bin,
a fluctuation in $NH_{Gal}$ (which, we remind, is obtained from an HI map smoothed on a scale of $\sim 1$ deg)
may bias one or both the temperature components of our model.  To investigate the
robustness of our strategy, we compare the best-fit values of $NH_{Gal}$, averaged across the
annuli in each cluster, with the value from \citet{2005Kalberla}.  As shown in Figure \ref{nh}, the values
are often consistent within 1$\sigma$, but a slight positive bias is observed with the best fit values being
on average $\sim 20$\% higher. We repeated our fit forcing $NH_{Gal}$ to be equal to the HI value, and
found that our results do not change qualitatively, except in the remarkable case of 2A0335, as
discussed in Section 3.  However, in many cases the quality of the fit is significantly worse,
and this is reflected in a systematic underestimation of the statistical error bars on the abundance values.
The results presented in this work therefore are conservatively based on the fits obtained with free $NH_{Gal}$.

\begin{figure*}
\begin{center}
\includegraphics[width=3.4in, height=3.2in, trim=100 210 100 200, clip]{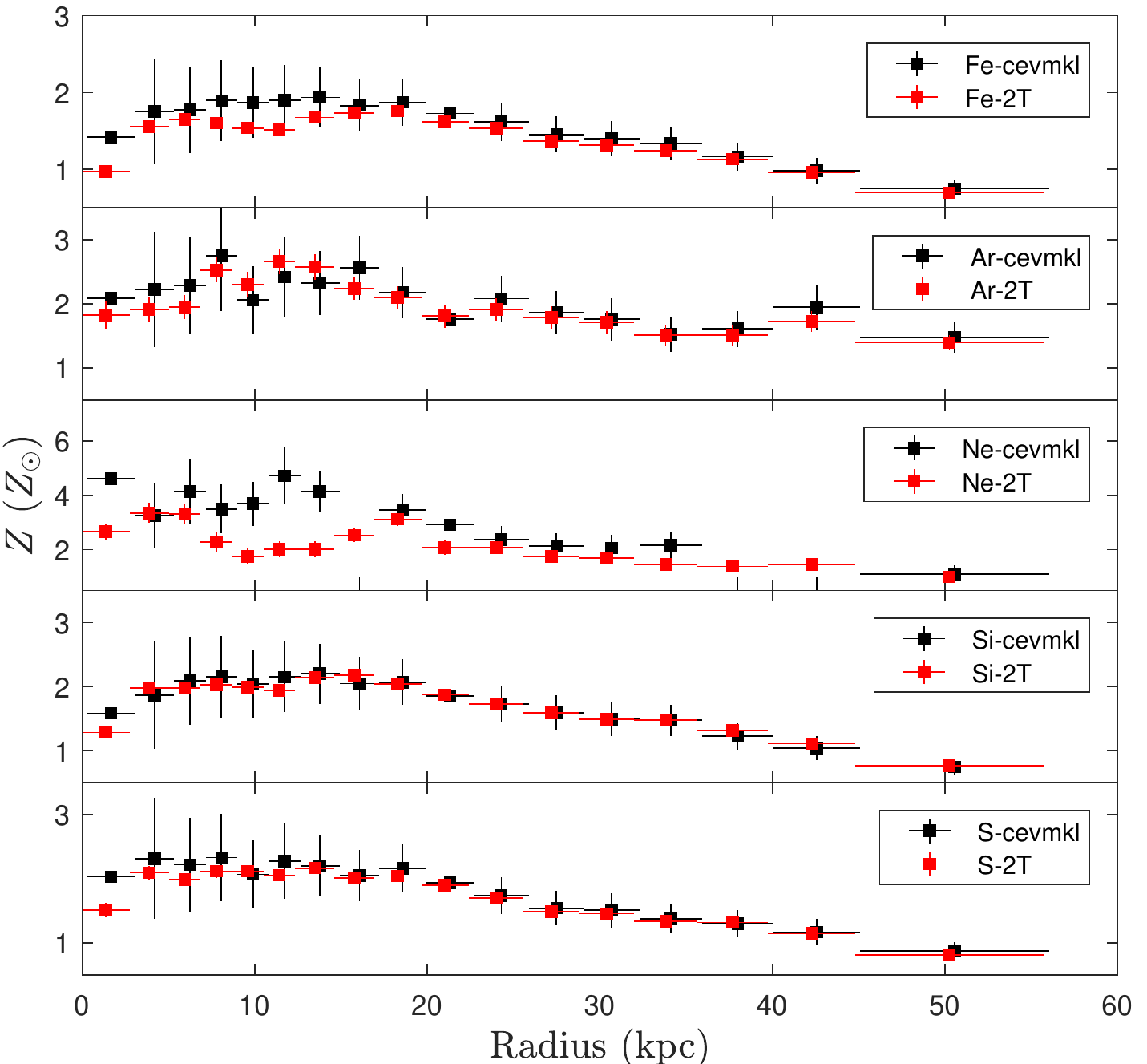}
\includegraphics[width=3.4in, height=3.2in, trim=80 210 100 200, clip]{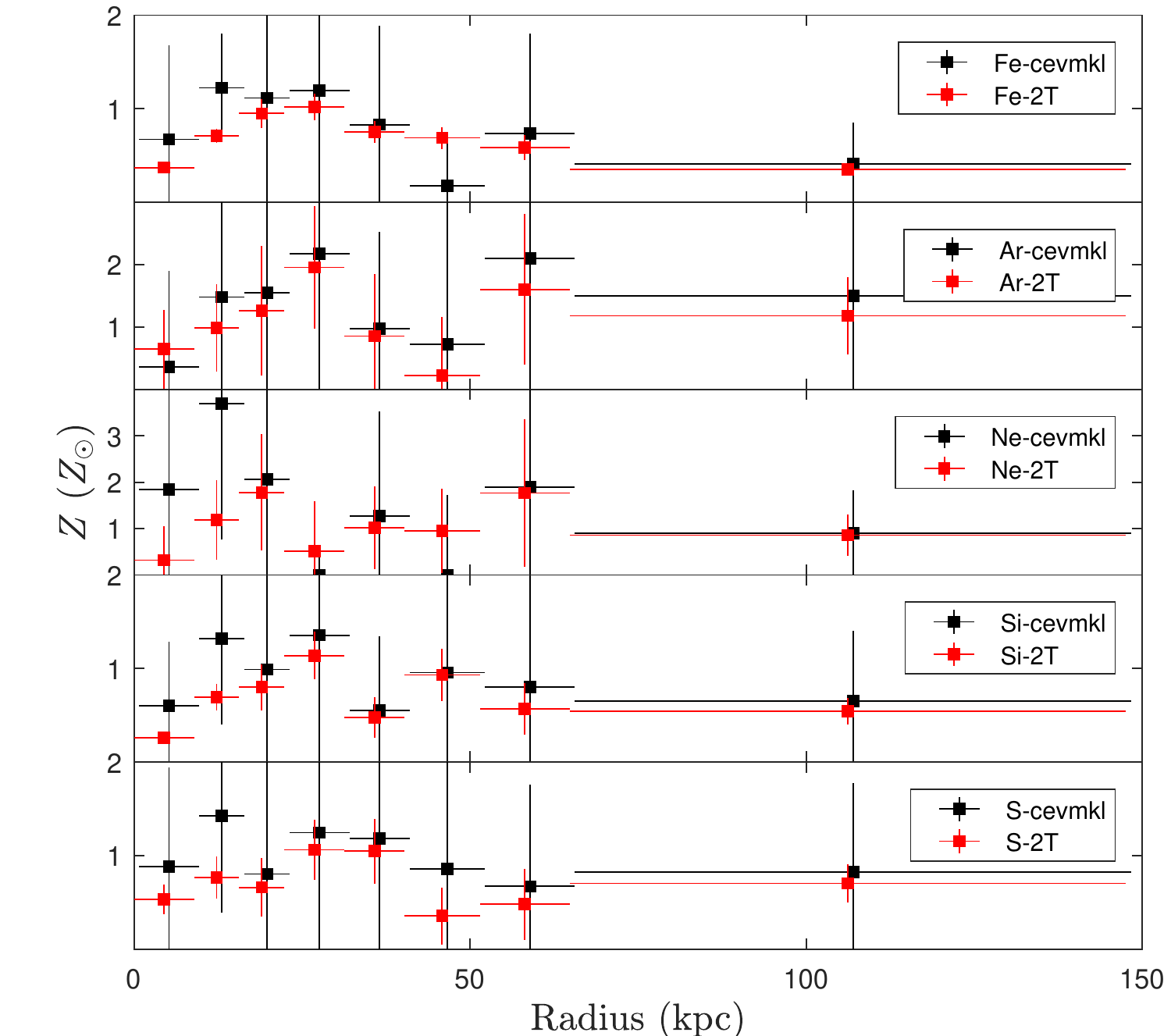}
\caption{The abundance profiles of Fe, Ar, Ne, Si and S for the  Centaurus (left panels) and Abell 1991
(right panels) obtained with the {\tt cevmkl} model (filled black squares) and our reference
two-temperature {\tt vapec} modelization (filled red squares). The points of the {\tt cevmkl} model have been
slightly shifted along the x-axis for clarity.}
\label{cevmkl}
\end{center}
\end{figure*}

A more fundamental and critical aspect of our analysis is the model of thermal structure of the ICM,
which we approximate with two temperature components.  We explore the effects of assuming a more complex
underlying temperature structure, despite it cannot be constrained efficiently with CCD data.  To do that,
we repeat our fit with a {\tt cevmkl} model, which is used to model the emission from a multi-temperature
plasma. Emission measures from different temperature components follow a power-law in temperature, while
the abundance ratios are defined in the usual way and are shared by all the temperature components. While we
fix the minimum temperature to a reference value of 0.5 keV, the maximum temperature is set to a value
somewhat larger than the largest temperature we found in our reference analysis, and leave the slope of
the emission measure-temperature relation free to vary.  In this way we can explore the effects of having
a continuous distribution of temperatures with respect to assuming only two values. We find that our results do
not change significantly, as shown in Figure \ref{cevmkl}, where we plot the Fe, Ar, Ne, Si and S
profiles obtained in the two modelizations (2T {\tt vapec} compared to {\tt cevmkl}) for Centaurus and
Abell 1991, which represent the highest and lowest quality spectra in our sample.
Not unexpectedly, the Ne profile is the only one that is noticeably affected, confirming our caveats that
our results for Ne strongly depend on the modelization of the $\sim 1$ keV line emission blend, and
therefore are particularly sensitive to the thermal model we assume.  Since we have no reasons to prefer a
{\tt cevmkl} rather than a double {\tt vapec} model, we decide to keep Ne profiles according to our
reference analysis, confirming the warning on unknown systematic uncertainties in addition to the
statistical errors. Finally, we warn that differences between the results from {\tt vapec} and
{\tt cevmkl} may also be due to different modelization of the Fe-L line complex built-in the two models.
In particular, our reference model {\tt vapec} is continuously updated, while {\tt cevmkl} and {\tt mekal}
are no longer updated since several years.  Some differences in the model may show up despite the
low resolution of CCD spectra, but are typically smaller than the typical statistical uncertainties we have
in our data.  A systematic comparison of the different atomic models goes beyond the goals of this paper.
\footnote{A detailed comparison can be found in {\tt http://www.atomdb.org/Issues/mekalspex.php}.}

Another source of potential systematics in our approach can be due to the use of a single abundance value common
to the two temperature components.  In general, the two temperatures should be considered a proxy for the real
temperature structure of the ICM along the line of sight.  This, in turn, can be due to the projection of ICM at
different temperature, or to the presence of different phases at the same radius, or both.  In any case, the
abundance values associated to different temperature components may well be different, and there are no compelling
reasons to force them to be equal.  This is expected on the basis of the much scattered but clear
correlation between metallicity and temperature in the core of clusters, or between metallicity and entropy
\citep[see][for a recent measurement of the abundance-entropy correlation in a sample of {\sl Chandra} clusters]{liu2018}.
However, because of the modest CCD spectral resolution, if we relax the hypothesis of a single abundance value
dominating the emission in each ring, the degeneracy between the temperature and abundance would be so large
to hamper the measurement of abundance radial profiles other then Fe.  The next major step forward to take
in this direction will be offered by X-ray bolometers, which will allow one to resolve single lines across the
soft X-ray spectrum and provide a mean to associate an abundance value to each temperature component,
enabling at the same time a much more structured thermal modelization.  Admittedly, in this work we
are compelled to assume a single abundance value for all the temperature components in each annulus, but
we are aware that this assumption may be tested at least in the only available data obtained with an
X-ray bolometer thanks to the {\sl Hitomi} mission \citep{hitomi2016nat}.  Therefore, we postpone this test to a
more detailed work on the Perseus cluster.

\begin{figure*}
\begin{center}
\includegraphics[width=0.32\textwidth, trim=115 258 150 273, clip]{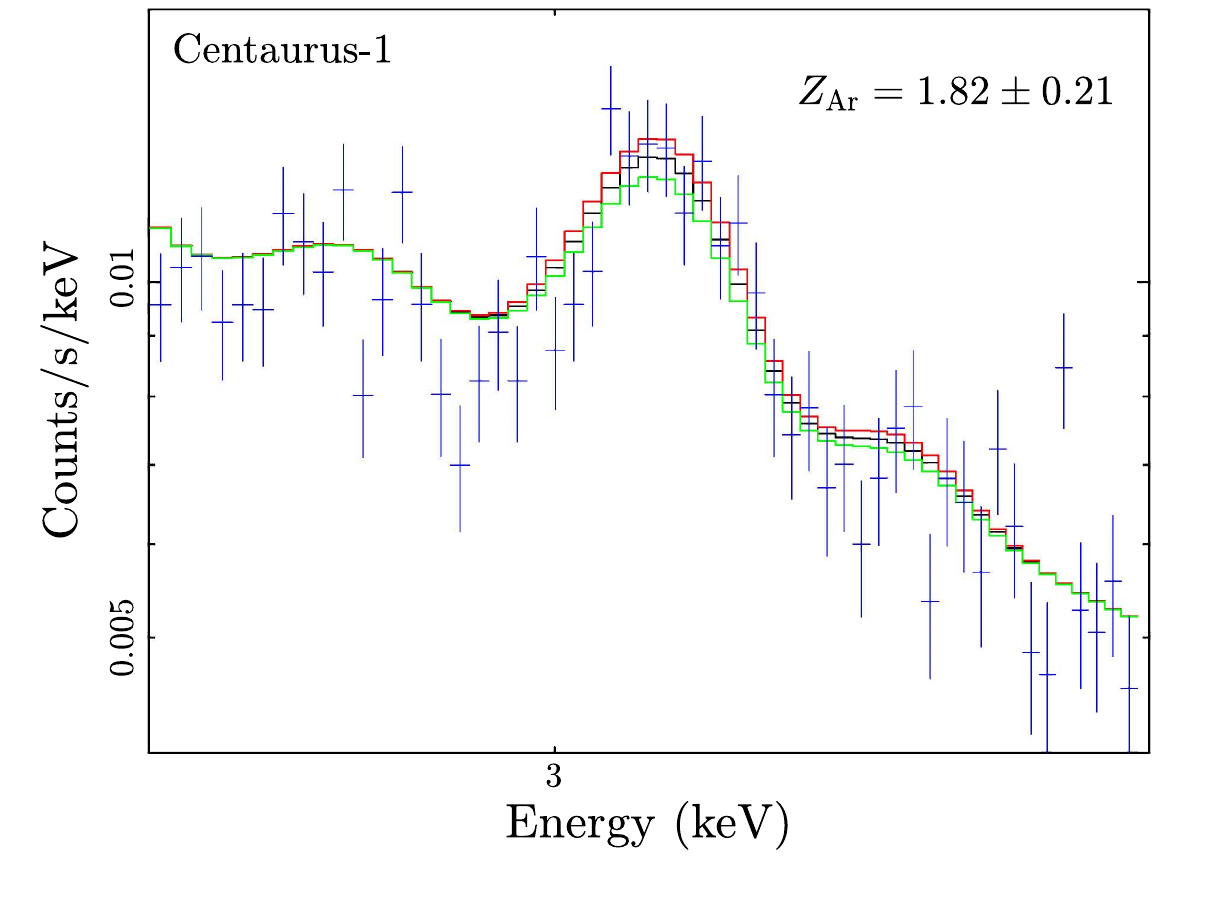}
\includegraphics[width=0.32\textwidth, trim=115 258 150 273, clip]{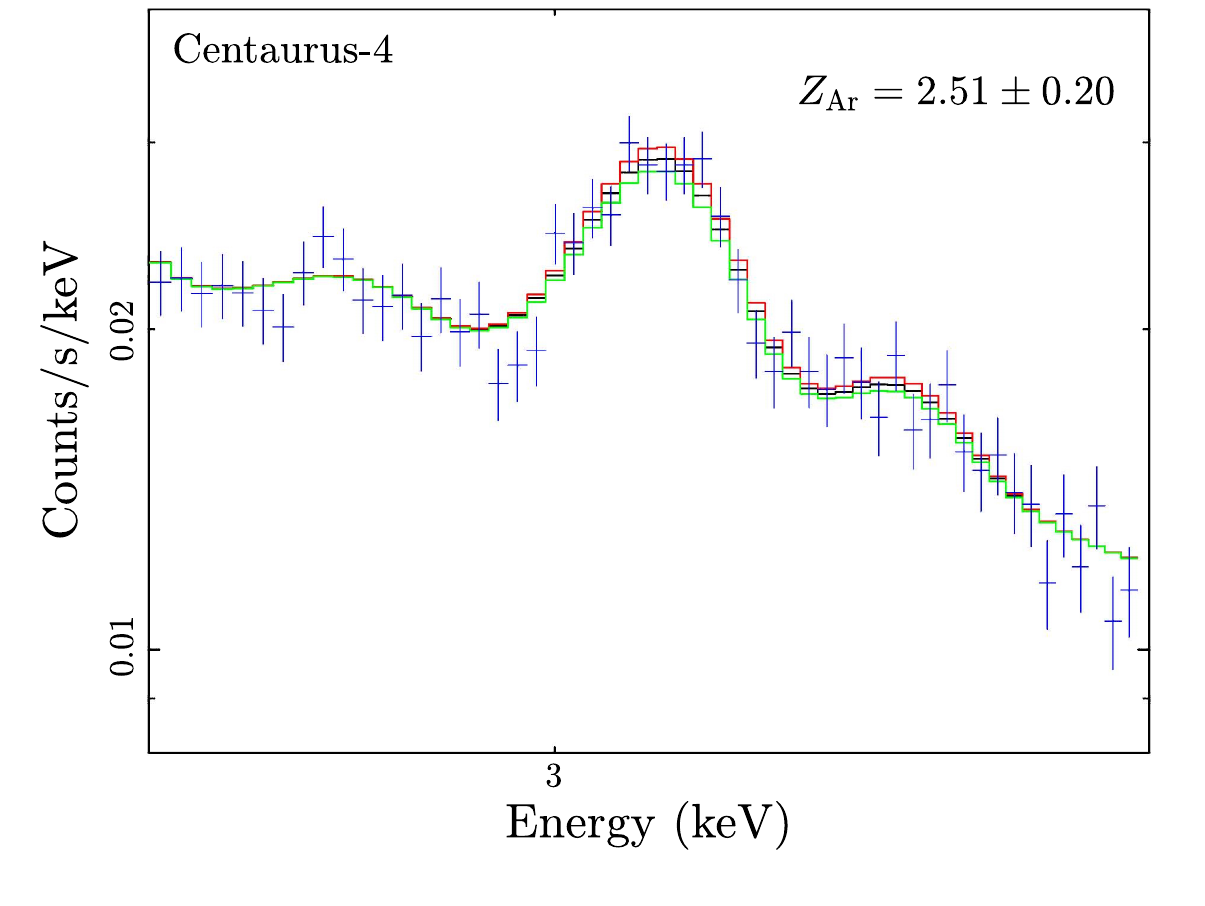}
\includegraphics[width=0.32\textwidth, trim=115 258 150 273, clip]{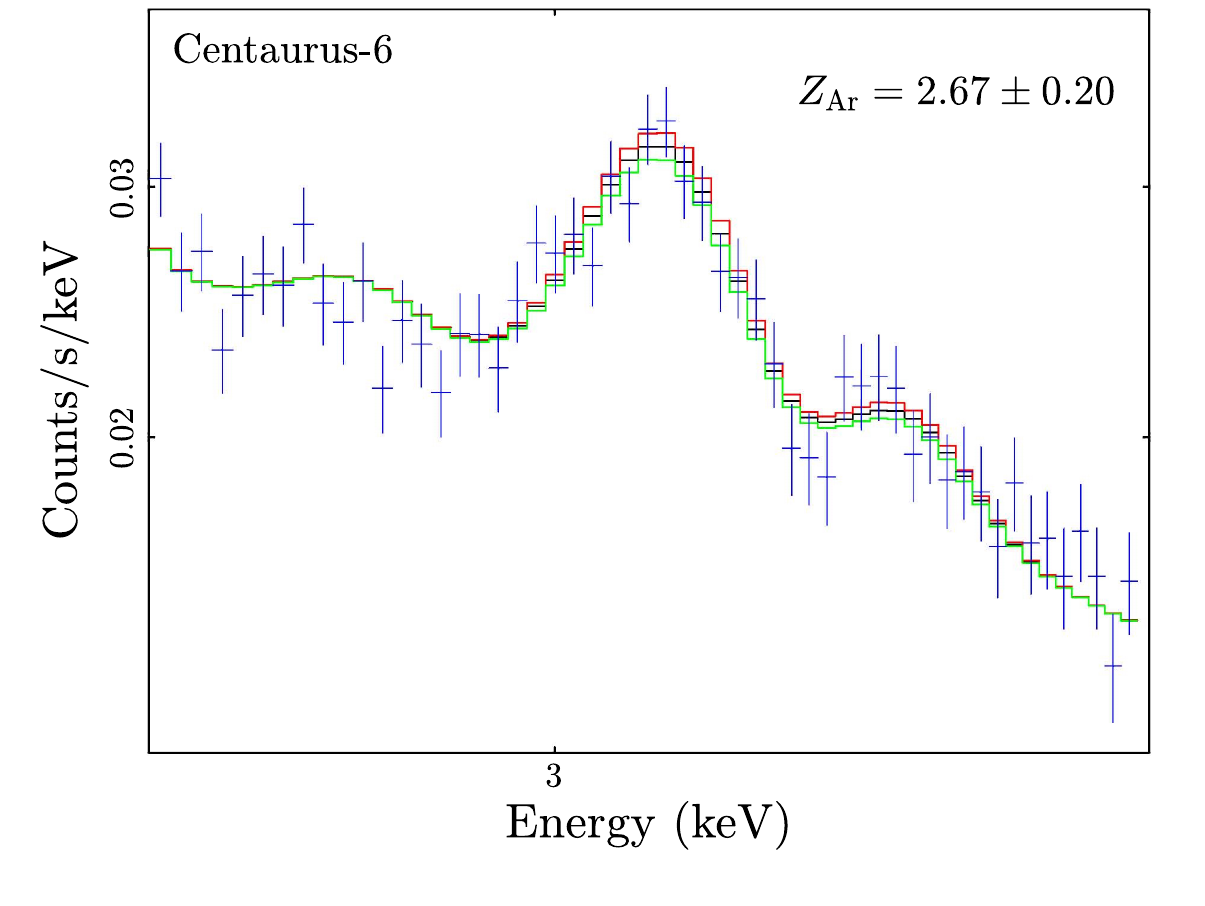}
\includegraphics[width=0.32\textwidth, trim=115 258 150 273, clip]{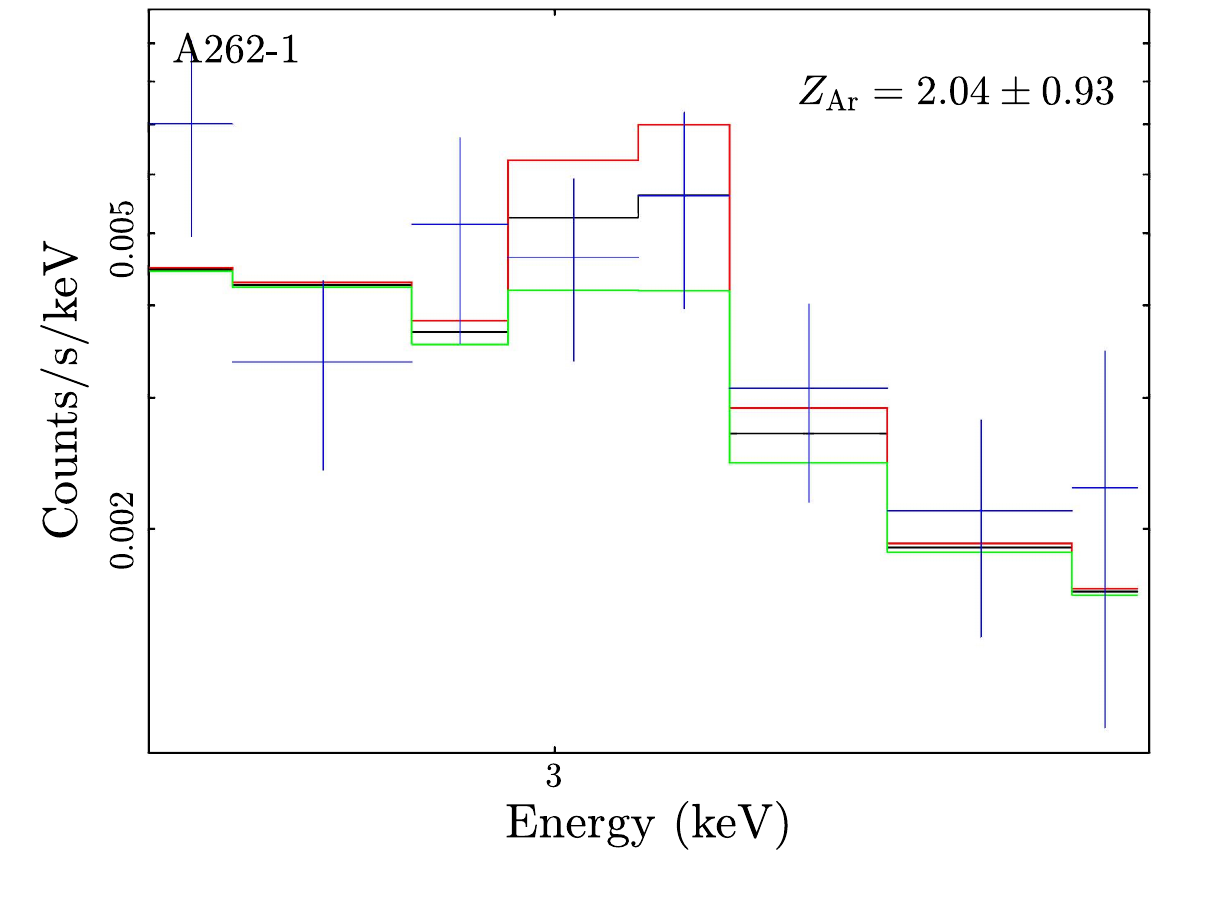}
\includegraphics[width=0.32\textwidth, trim=115 258 150 273, clip]{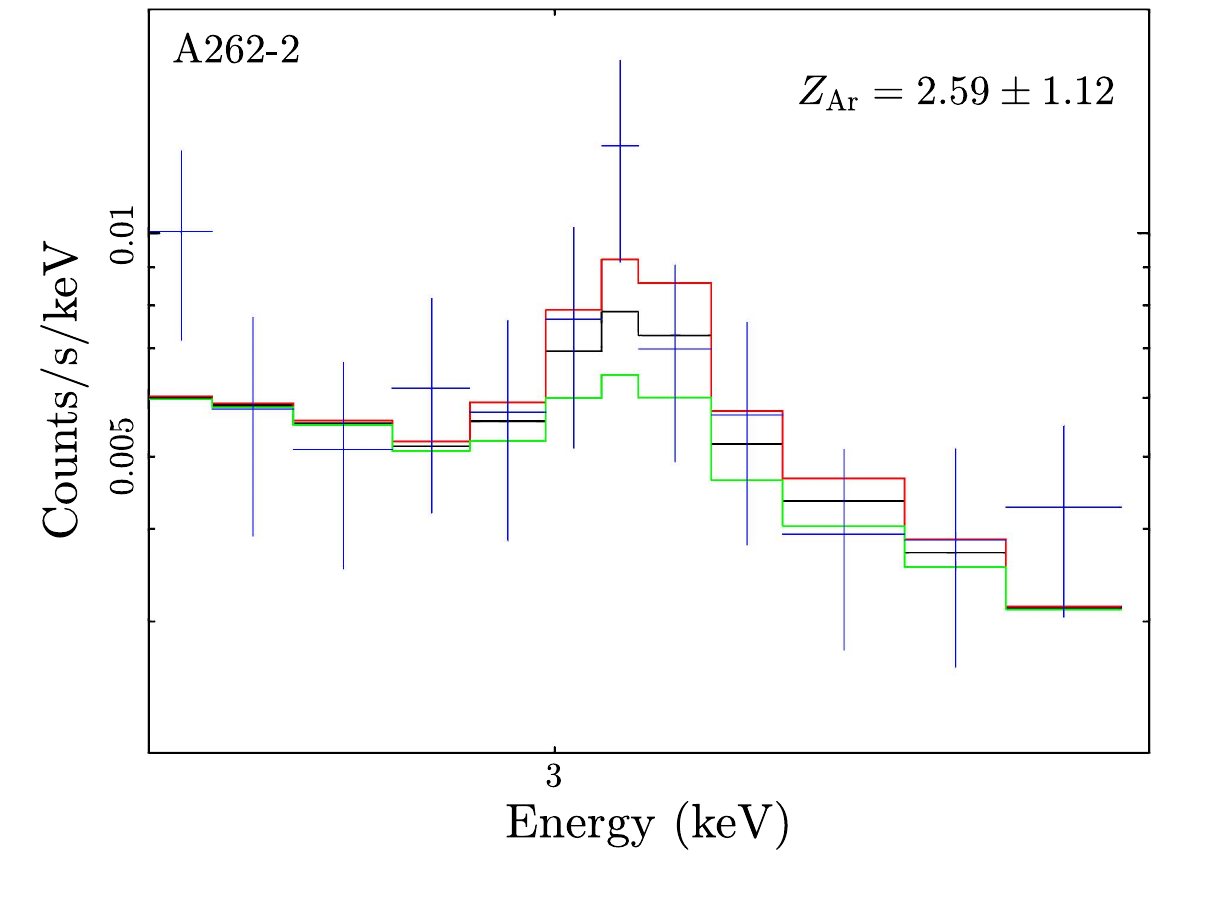}
\includegraphics[width=0.32\textwidth, trim=115 258 150 273, clip]{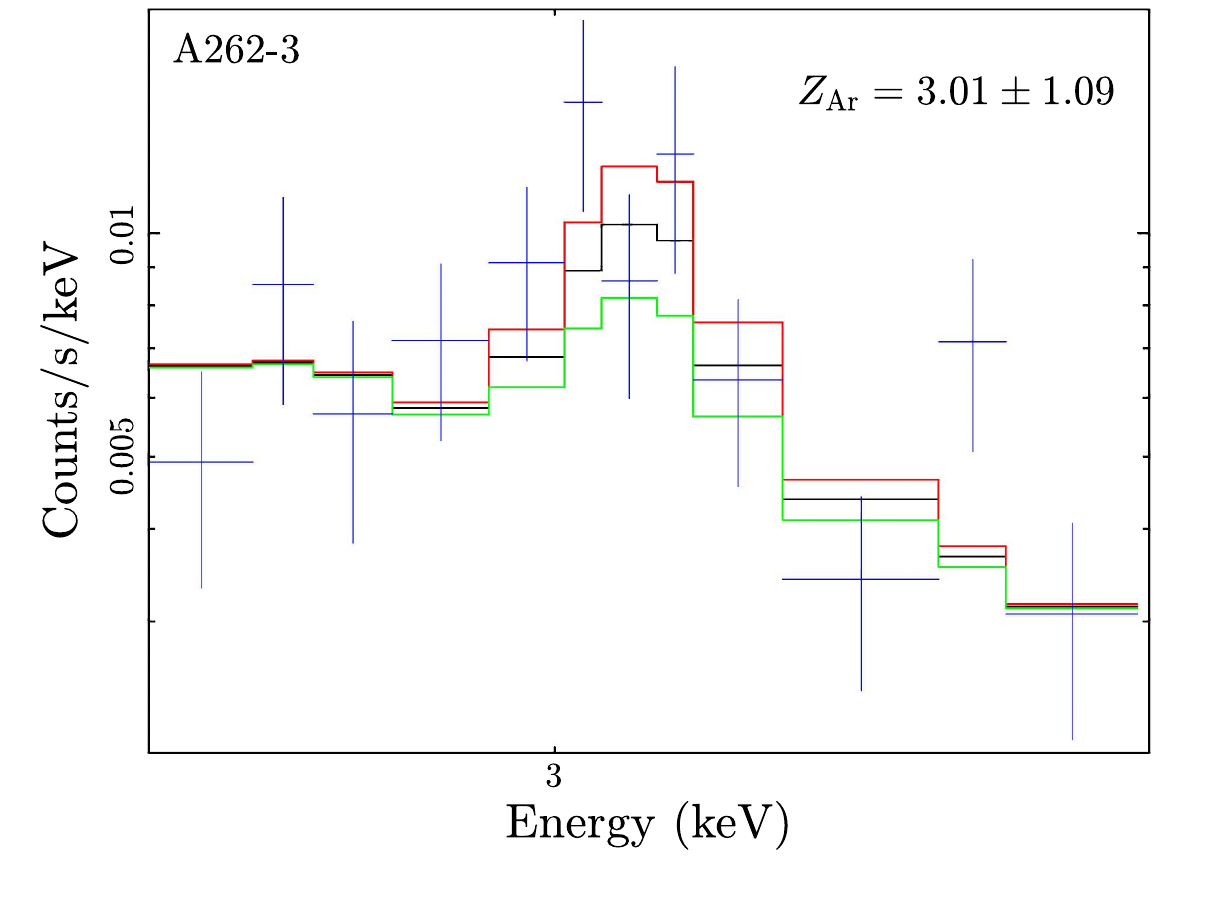}
\includegraphics[width=0.32\textwidth, trim=115 258 150 273, clip]{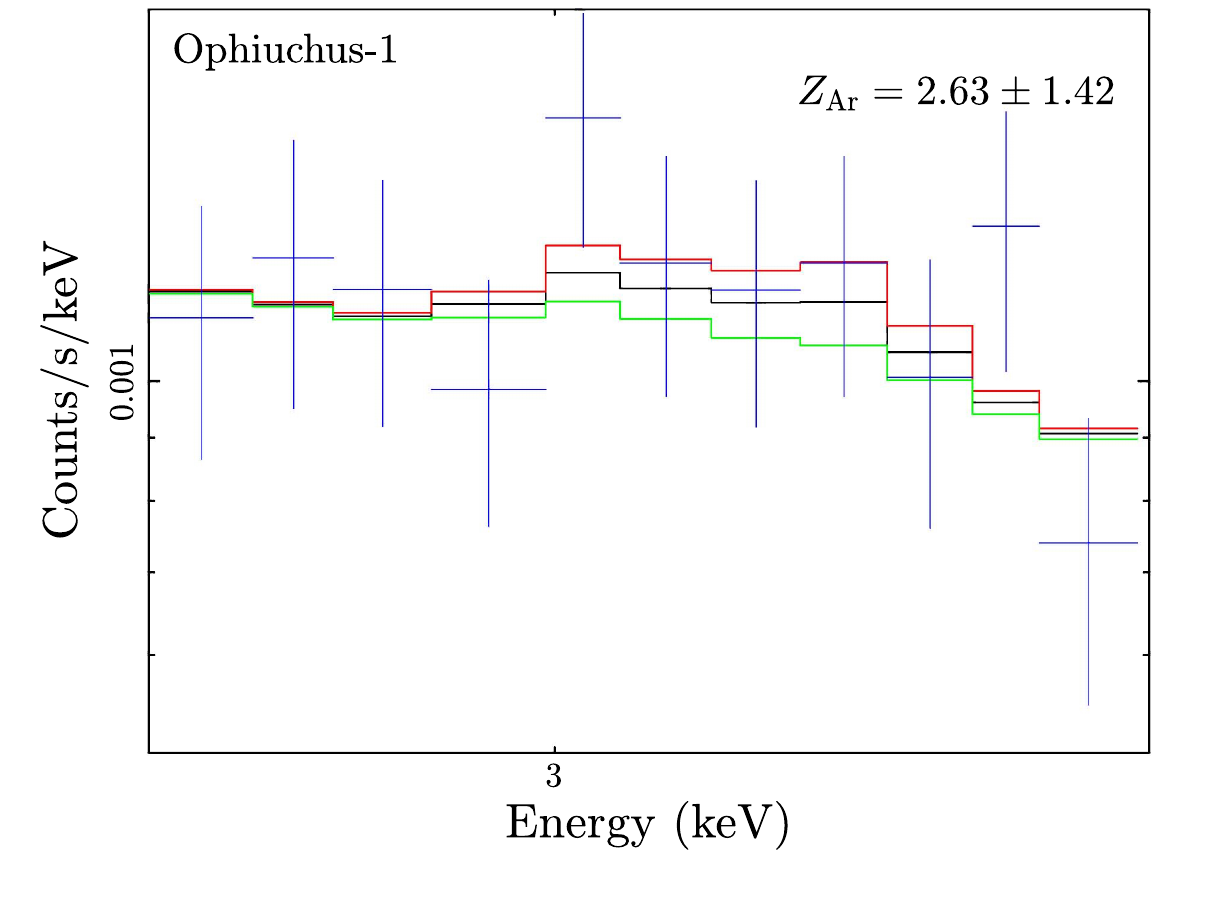}
\includegraphics[width=0.32\textwidth, trim=115 258 150 273, clip]{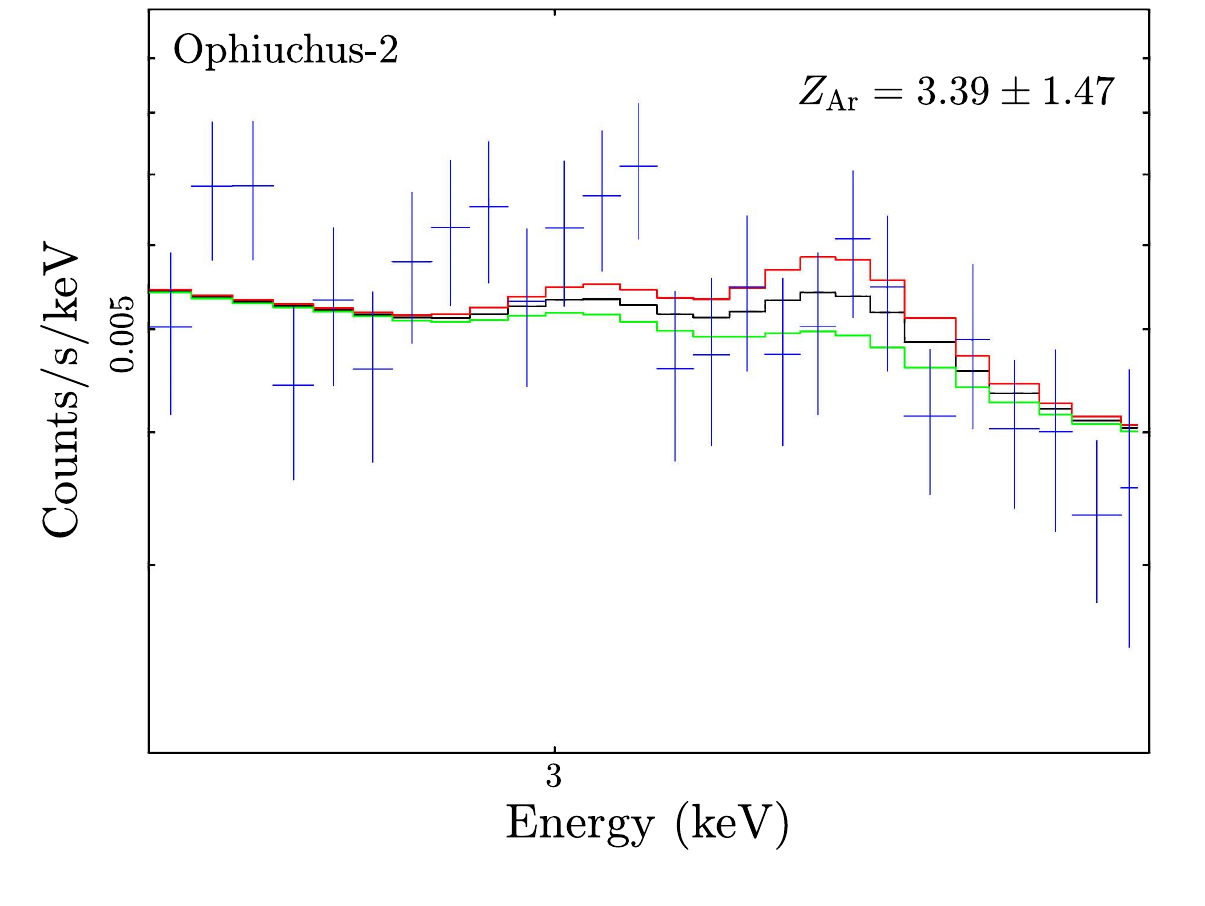}
\includegraphics[width=0.32\textwidth, trim=115 258 150 273, clip]{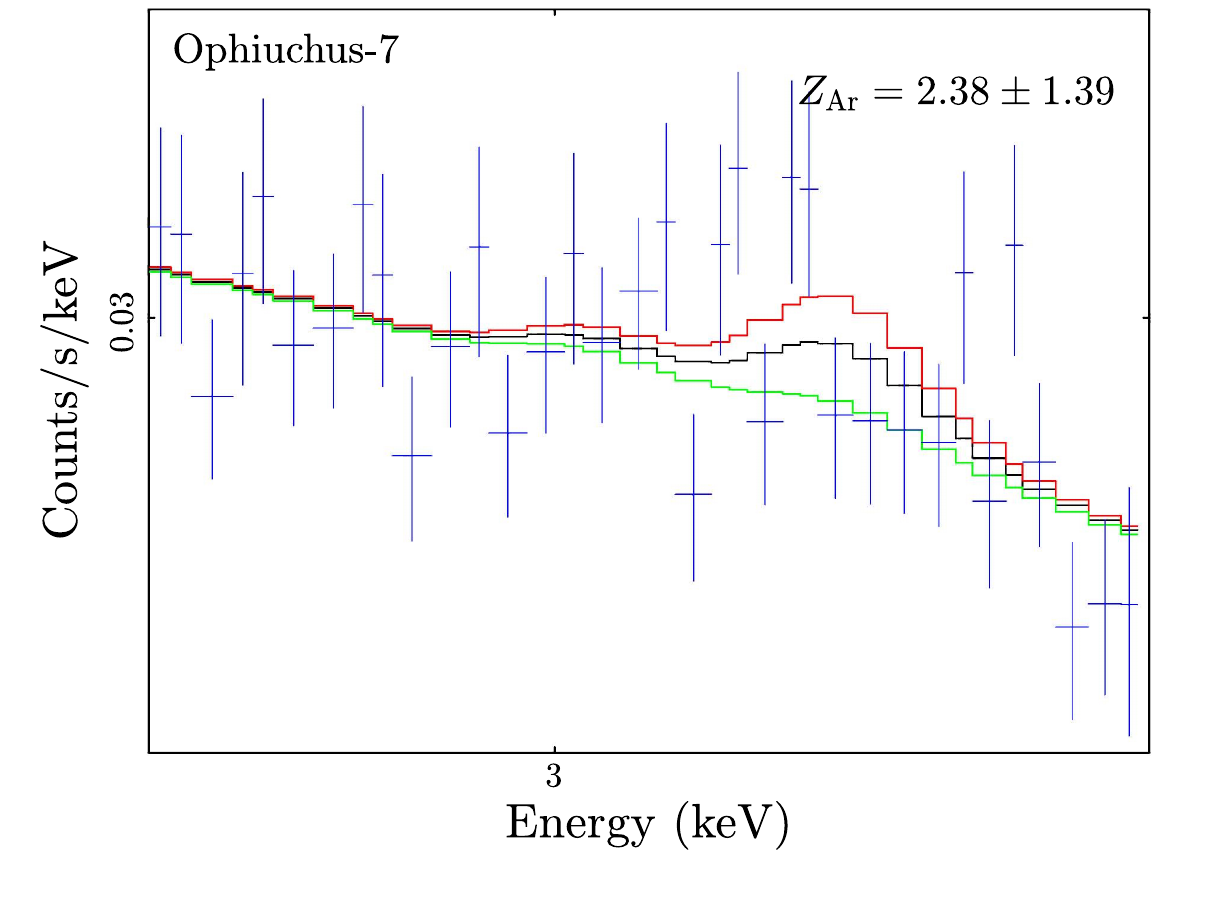}
\caption{Narrow band (2.7--3.5 keV) stacked spectra of the first, fourth, and sixth bins of Centaurus, the
first, second, and third bins of A262, and the first, second, and seventh bins of Ophiuchus. The
best-fit models with 1$\sigma$ upper and lower envelopes,
as obtained from the broad-band fit, are shown as black, red and green solid lines, respectively.}
\label{spec}
\end{center}
\end{figure*}

\begin{figure}
\begin{center}
\includegraphics[width=0.49\textwidth, trim=82 220 77 240, clip]{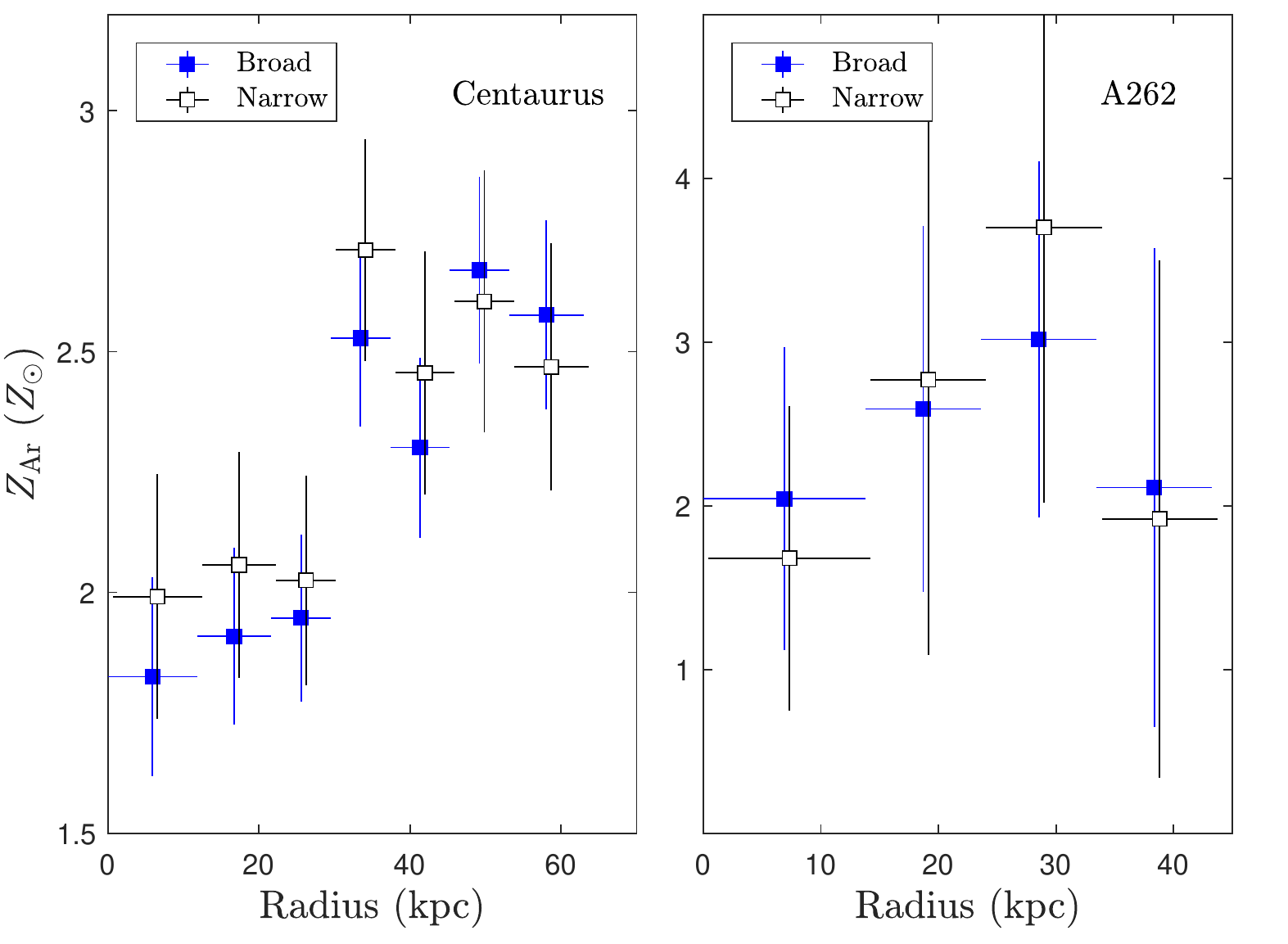}
\caption{$Z_{\rm Ar}$ profiles of the inner most regions of Centaurus and A262. Solid-blue and empty-black squares represent the results from the 0.5--7 keV broad band, and the 2.9--3.5 keV narrow band, respectively. The black data points have been slightly shifted along the x-axis for clarity.}
\label{narrow}
\end{center}
\end{figure}

Finally, it has been found that effective area calibrations are different among major, currently-available
X-ray CCDs \citep[see][]{simionescu2019}, and this effect can bias the measurement of metal abundance,
particularly when the fit is driven by energy range far from that where the relevant lines are.  In other
words, the level of continuum can be poorly fitted, and this can severely affect the equivalent width of
faint lines.  For example, we notice that in a few cases (Centaurus, A262 and Ophiuchus) the maximum measured
value of the Ar abundance is remarkably higher than $2\, Z_\odot$.  To investigate whether these high values
were actually spurious, first we perform a visual check, finding that the continuum level appears to be
properly fitted in our spectra.  A few relevant examples are presented in Figure \ref{spec}, where we show
the 2.7--3.5 keV range of the stacked spectra corresponding to three bins of Centaurus (bin 1 , 4, and 6 with
$Z_{\rm Ar}=1.82\pm0.21$, $2.51\pm0.20$ and  $2.67\pm0.20$, respectively), A262 (bin 1, 2, and 3 with
$Z_{\rm Ar}=2.04\pm0.93$, $2.59\pm1.12$ and $3.01\pm1.09$, respectively), and Ophiuchus (bin 1, 2, and 7
with $Z_{\rm Ar}=2.63\pm 1.42$, $3.39\pm 1.47$ and $2.38\pm 1.39$, respectively).  We recall that the
best-fit models, shown as solid lines, are obtained separately fitting the spectra from single Obsid, and
not the stacked spectrum.  In all the cases the uncertainties in the continuum appear to be well within
the statistical uncertainties, despite for the Ophiuchus cluster the spectrum is very noisy and the
uncertainty on Ar abundance may be larger than the estimated error bars.  We verified that our results do
not change after removing the profile of the Ophiuchus cluster from the stacked abundance profiles.
We conclude that the unexpectedly high values of Ar abundance found in some bins are properly characterized
and have well-measured uncertainties, except one case that may be affected by systematics, but is not relevant
to our main conclusions.

To better test the robustness of our results with respect
to calibration uncertainties, we adopt the strategy followed in \citet{lakhchaura2018}, where the
abundances are remeasured in the 2.9--3.5 keV narrow band by leaving the normalization free, and the
temperature, Galactic absorption frozen to the best-fit values of the broad band fit. In particular, we
show in Figure \ref{narrow} the Ar abundance profiles in Centaurus and A262 computed in the
narrow band and broad band fits. Although some difference is present, the significance of the drop in the
Ar abundance is not affected. Therefore, we confirm
that our results are not significantly affected by clear uncertainties in the broad-band Chandra calibration.

\section{Conclusions}

In this work we analyze the abundance profiles in clusters where a central drop
in the Fe abundance has been reported in the literature. Our aim is to test the
origin of this feature, which may be associated to a mechanical process, like the uplifting
of highly-enriched ICM in the center by AGN feedback, or to Fe depletion into dust grains, or
to both processes.
We exploit the exquisite angular resolution of {\sl Chandra} to measure the abundance in the
innermost regions of a sample of 12 cool-core clusters, thanks to the spectral resolution of the ACIS detectors.
In particular, we compute the abundance profile of Ar and, within the limits of our assumptions, Ne,
that cannot be incorporated into dust, and therefore should not exhibit a central abundance drop
if due to dust depletion, when compared to the abundance profiles of other elements like
Fe, Si or S.  Our conclusions are summarized as follows.

\begin{itemize}

\item We confirm the detection of the Fe drop in 10 out of 12 clusters at a significance level
larger than 2$\sigma$, in broad agreement with the literature.

\item We are able to compute Ar abundance profiles out to $\sim0.15r_{500}$
after assuming a two-temperature structure for the ICM.  We also present the profiles of
Ne, which, however, are likely affected by systematics due to partial modeling of the
line emission blend including Ne lines.  We are, however, able to show that our abundance
profiles for all the other metals (Fe, Ar, Si and S) are not significantly affected by our
specific choice for the thermal modelization of the ICM.

\item We find central abundance drops of Ar in 4 clusters at a confidence level of
more than 2$\sigma$, and, formally, 4 clusters with abundance drops of Ne.
Si and S abundance profiles are found to be broadly consistent with Fe.
Overall, our results are consistent with all the metals showing a central drop in abundance,
therefore suggesting that a mechanical process directly removing the highly enriched
ICM from the center, as the observed uplift of dense ICM driven by cavities carved by the central AGN,
should be the main origin of the central abundance drops.

\item We further extend our results by stacking the profiles in six bins in the range [0--0.15]$r_{500}$,
finding that Ar (and possibly Ne) has significantly larger abundance in the center with respect to
Fe, Si and S, and that its profile shows a much steeper gradient.  This result, in turn,
suggests that the dust grain scenario is indeed taking place in the center of these clusters,
and has a significant effect in producing the Fe drop, which sums up to the mechanical removal
of the most enriched gas. We successfully test the robustness of our conclusions with respect to the
thermal modelization of the ICM, the spectral analysis strategy and possible calibration uncertainties.

\item We confirm the detection of central Fe abundance drop in the galaxy cluster MACSJ1423.8+2404 at redshift
0.543, indicating that this feature is not confined to low redshift clusters, but has occurred
in this cool-core cluster at least $\sim6$~Gyr ago.

\end{itemize}

To summarize, we conclude that the signature of dust depletion can be observed
in the abundance profiles of nearby, bright groups and clusters as a difference between the abundance profiles
of noble gas like Ar and Ne and that of dust-depletable elements like Fe, Si and S.
While some additional work can be done in this field on the basis of CCD data from the {\sl Chandra}
and XMM-Newton archives, a major improvement
will be enabled only by the use of future X-ray bolometers.  The capability of measuring with high accuracy
the amount of Fe lost from the ICM because of dust depletion will be an important information to constrain
the cycle of baryons in the center of cool-core clusters and the interplay of the ICM with
the insterstellar medium of the BCG.

\section*{acknowledgements}

We thank Simone Bianchi for useful discussions. We also thank the anonymous referee for
a detailed and constructive report which significantly improved the paper.  We acknowledge
financial contribution from the INAF PRIN-SKA 2017 program 1.05.01.88.04 ({\sl ESKAPE}) and
from the agreement ASI-INAF n.2017-14-H.O.

\bibliography{iron_drop}

\end{document}